\documentclass{article}

\usepackage{times}  
\usepackage{titlesec}
\usepackage{hyperref}
\usepackage{tikz}
\usepackage{amsmath, amssymb}
\usepackage{graphicx}
\usepackage{booktabs,tabularx,threeparttable,float}
\usepackage{xspace}
\usepackage{xcolor}
\usepackage[T1]{fontenc}
\usepackage{subfigure}
\usepackage{subcaption}
\usepackage{verbatim}
\usepackage{url}
\usepackage{enumitem}
\usepackage{listings}
\usepackage{multirow}
\usepackage{pifont}
\usepackage{mathtools}
\usepackage{fancyhdr}
\usepackage{caption}
\usepackage{xurl}
\usepackage{longtable}

\newcommand{\sysname}{\texttt{SIMPLE}\xspace}
\newcommand{\sysnames}{\texttt{SIMPLE}’s \,}

\newcommand{\para}[1]{\smallskip\noindent\textbf{#1}}

\newcommand{\num}[1]{\normalsize{\textcircled{\scriptsize{#1}}}\normalsize\xspace}

\newcommand{\UseDisplaySpacing}{%
}

\usepackage{microtype}

\usepackage{hyperref}

\usepackage[accepted]{mlsys2025}

\mlsystitlerunning{SIMPLE: Disaggregating Sampling from GPU Inference into a Decision Plane for Faster Distributed LLM Serving}

\begin{document}

\twocolumn[
\mlsystitle{SIMPLE: Disaggregating Sampling from GPU Inference into a Decision Plane for Faster Distributed LLM Serving}




\mlsyssetsymbol{equal}{*}

\begin{mlsysauthorlist}
\mlsysauthor{Bohan Zhao}{scitix}
\mlsysauthor{Zane Cao}{scitix}
\mlsysauthor{Yongchao He}{scitix}
\end{mlsysauthorlist}

\mlsysaffiliation{scitix}{SCITIX (SGP) TECH PTE. LTD.}

\mlsyscorrespondingauthor{Bohan Zhao}{bhzhao@scitix.ai}

\mlsyskeywords{Machine Learning, MLSys}

\vskip 0.3in

\begin{abstract}
As large language models (LLMs) scale out with tensor parallelism (TP) and pipeline parallelism (PP) and production stacks have aggressively optimized the \emph{data plane} (attention/GEMM and KV cache), \emph{sampling}—the \emph{decision plane} that turns logits into tokens—becomes a new bottleneck. This creates a structural \emph{holdout}: sampling neither expands with TP nor balances across PP stages, so its share of iteration time \emph{grows} as GPUs get faster and it caps pipeline frequency at the last stage. We present \sysname, a stage-agnostic, sequence-parallel, overlappable decision plane that disaggregates sampling into a CPU-side service and shrinks its runtime footprint back to a minor, hidden role. \sysname combines: (1) \emph{sequence-parallel sampling}, which shards work along the batch dimension and removes vocabulary-axis collectives; (2) a CPU-based algorithm with \emph{column-wise penalties} and \emph{truncation-first} filtering to realize single-pass, linear-time kernels; and (3) \emph{speculative hot-vocab sampling} (SHVS), which samples on a small hot set with rejection-correctness and uses a simple sizing model to choose the hot-vocab size that maximizes throughput. In evaluation, \sysname improves end-to-end throughput by up to 96\% and reduces P95 latency by 20–65\%. Crucially, \sysname requires no user-side code changes and composes with existing data-plane optimizations, unlocking scaling benefits that compound with future GPU generations.

\end{abstract}
]

\section{Introduction}

In modern online serving, the parameter size of mainstream LLMs~\cite{ouyang2022gpt,gptoss,liu2024deepseekv3,team2025kimik2, qwen2.5, brown2020gpt3} typically exceeds the memory capacity of a single GPU. Even high-end accelerators such as the H100 (80\,GB), H200 (141\,GB), and B200 (192\,GB) provide insufficient memory, while earlier generations offer much less. Model weights alone can occupy several hundred gigabytes—for instance, DeepSeek-R1 and Qwen3-235B-A22 require about 670\,GB and 470\,GB, respectively—excluding the additional space reserved for key-value (KV) caches. As a result, production inference commonly spans multiple GPUs~\cite{vllm_ds3,lambda_llama405,google_serve}, often across hosts via tensor (TP), pipeline (PP), expert parallelism (EP), or their combinations~\cite{shoeybi2019megatron,guo2025deepseek}.

Modern LLM inference proceeds in two successive steps per \emph{iteration} (\S\ref{subsec:autoregressive}): 
\emph{(i) Forward.} The GPU \emph{data plane} executes attention and feed-forward kernels, moves/updates KV-cache, and produces \emph{logits}. 
\emph{(ii) Sampling.} The \emph{decision plane} selects the next token by sampling from the probability distribution of logits over the model’s vocabulary.

Existing efforts have primarily optimized the \emph{data plane}—accelerating matrix multiplications~\cite{DeepSeek2025DeepGEMM,Elhoushi2025Any4,Lin2025QServe} and improving KV-cache transport/placement~\cite{li2024snapkv,cai2024pyramidkv,kwon2023vllm,liu2024cachegen,xiao2023sink,zhang2023h2o}. Together with faster GPUs, these advances have shortened compute and communication in the forward step. However, they leave the \emph{decision plane} (sampling) largely unchanged, where sampling remains a serial epilogue at the end of each iteration. Meanwhile, vocabulary sizes have expanded markedly (\S\ref{sec:bg_vocab}), increasing the memory-bound computation per sampling step. Consequently, sampling now occupies a growing fraction of iteration time (\S\ref{sec:motivation}), emerging as a structural bottleneck in large-scale, model-parallel inference.


\para{Sampling is non-negligible.}
Contrary to common assumptions, sampling is parallel-unfriendly and can become a bottleneck in distributed settings for two reasons.
(i) \emph{Not TP-expandable.} TP shards the \emph{hidden} dimension of logits (later projected to the vocabulary) across ranks, whereas the heavy parts of sampling are \emph{vocabulary-axis} operations. Forming a \emph{global} decision therefore requires reconciling shard-local views (e.g., all-gathering logits), which effectively halts parallelization early. As we scale out with more ranks, the non-parallelizable sampling fraction grows, as suggested by \emph{Amdahl's law}~\cite{amdahl1967validity}. Consequently, the serial epilogue dominates throughput scaling (\emph{Gustafson's law}~\cite{gustafson1988reevaluating}): empirically, sampling’s share reaches up to \(38\%\) on large-vocabulary models and rises by \(\sim\!10\%\) as tensor parallelism grows from 2 to 8.
(ii) \emph{Not PP-balanced.} Sampling is executed only at the \emph{last} PP stage, extending its stage time and capping the pipeline frequency at the stage maximum. This reintroduces bubbles (22–40\%) when compute stages are balanced; by \emph{Little’s law}~\cite{little1961proof}, the longer last stage inflates queueing and tail latency.
In short, the causes of sampling cost do not shrink with faster GEMMs, so the epilogue becomes the new critical path. It is therefore time to \emph{re-architect} sampling so it returns to an inconspicuous final step rather than an invisible performance hog.

\begin{figure}[t]
    \UseDisplaySpacing
  \centering
  \subfigure[Sampling ratio $f$ vs.\ TP degrees.\label{fig:sampling_ratio2}]{
    \includegraphics[width=0.9\columnwidth]{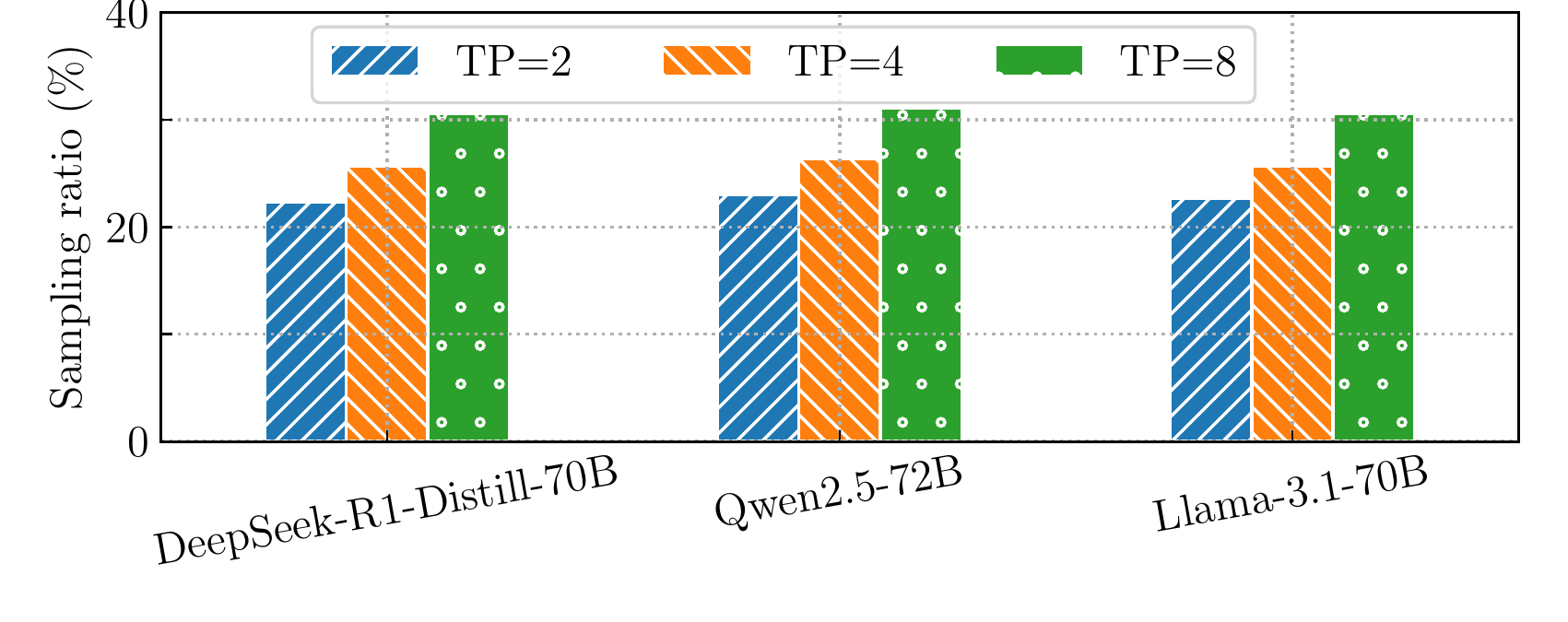}
  }

  \subfigure[Per-iter breakdown with Qwen-2.5-72B ($t{=}4$, $p{=}2$)]{%
    \includegraphics[width=0.9\columnwidth]{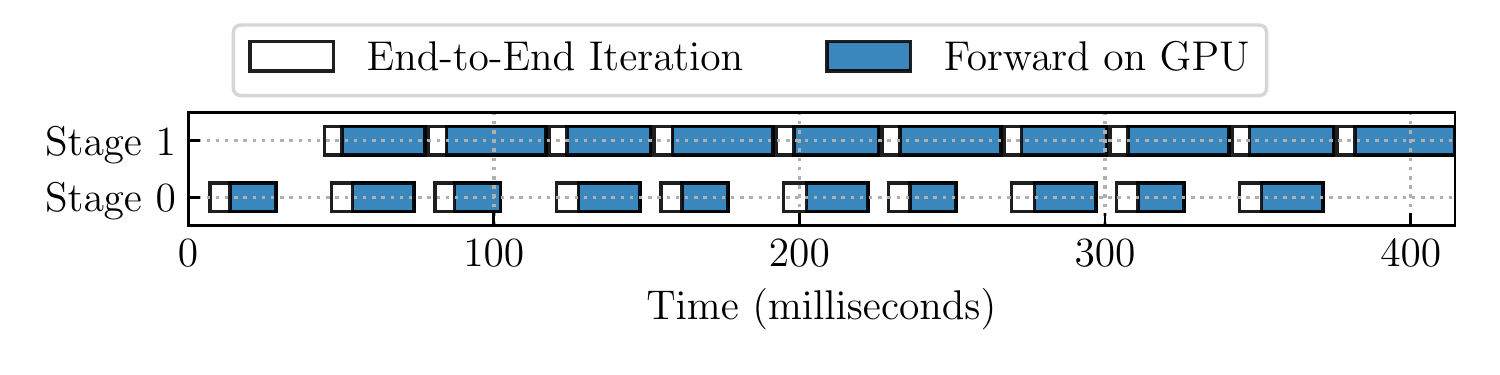}
    \label{fig:breakdown_iteration_qwen_72B}
  }

  \caption{
  Sampling bottlenecks in inference on 8$\times$H100 .Bars denote \emph{iteration time}; filled regions denote \emph{computation time}.
  }
  \label{fig:sampling_ratio_combined}
\end{figure}

\paragraph{A modern decision plane.}
To overcome the bottlenecks above, sampling should be \emph{disaggregated} and redesigned around three properties aligned with modern inference: it must be \emph{parallelizable} (split cleanly across TP members without re-materializing full logits), \emph{stage-agnostic} (kept off the PP critical path to avoid stage skew), and \emph{overlappable} (its latency hidden under GPU compute).

To this end, we present \sysname, a pluggable, standalone sampling service that attaches to existing engines with no user-side code changes. \sysname re-architects the decision plane with three complementary techniques that map one-to-one to the above goals: (1) \emph{sequence-parallel sampling}, which shards work along the batch (sequence) dimension to deliver TP-friendly parallelism without vocabulary-axis collectives (\textit{parallelizable}); (2) \emph{CPU offloading} with \emph{column-wise penalties} and \emph{truncation-first} passes to realize single-pass, linear-time kernels and decouple sampling from the last GPU stage (\textit{stage-agnostic}); and (3) \emph{speculative hot-vocab sampling}, which exploits Zipf distribution to replace the full-vocabulary scan with a fast path on the hottest sub-vocabulary corrected by rejection sampling, keeping the decision work beneath GPU compute in the common case (\textit{overlappable}). 

The contribution of this paper includes: 
(1) We identify and quantify sampling as a distributed-inference \emph{holdout}, showing that its share of end-to-end throughput and stall time \emph{increases} with larger TP and deeper PP, and we analyze why current designs fail to scale.
(2) We formulate sampling as a \emph{decision-plane service}, orthogonal to data-plane optimizations, that follows a \emph{speculate-then-correct} paradigm: sequence parallelism, CPU offloading with column-wise/truncation-first passes, and \emph{speculative hot-vocab} with rejection-correctness (distributionally exact).
(3) We design and integrate \sysname with existing stacks, improving end-to-end throughput by up to \textbf{96\%} and reducing P95 latency by up to \textbf{65\%}, while lowering GPU residency attributable to sampling—all with modest CPU assistance.
(4) We demonstrate substantial end-to-end gains—higher throughput and lower P50/P99 latency—and fewer pipeline bubbles across models and scales, while preserving output quality.

\section{Background}
\label{sec:background}

\subsection{LLM Inference}
\label{subsec:autoregressive}

We first outline the workflow of modern LLM inference: A pretrained model takes a textual \emph{prompt} as input and produces a sequence of \emph{output tokens}. These tokens are represented as discrete IDs through a tokenizer~\cite{gage1994new,sennrich-etal-2016-neural,kudo-richardson-2018-sentencepiece}, which maps token IDs to \emph{text} in a fixed \emph{vocabulary} \(\mathcal{V}\) of size \(V\). 

During inference, an \emph{engine}~\cite{zheng2024sglang,kwon2023vllm} maintains the state of each request as a \emph{sequence}—consisting of prompt and generated token IDs—and advances decoding in discrete \emph{iterations}. At iteration \(s\), each active sequence contributes exactly one new token. To utilize GPUs efficiently, sequences at the same iteration are grouped into a \emph{batch} and executed in parallel. 
Each iteration then proceeds in two steps:

\para{Forward.}
Regardless of whether PP or TP is employed, the \emph{forward} propagation eventually produces a matrix of \emph{logits} \(\mathbf{Z}_s \in \mathbb{R}^{B \times V}\), where \(B\) denotes the micro-batch size and \(V\) the vocabulary size.

\para{Sampling.}
Let \(\mathbf{Z}_s^{(b)}=\mathbf{Z}_s[b,:]\in\mathbb{R}^{V}\) be the logits row for sequence \(b \in \{1,\dots,B\}\) at iteration \(s\), and \(\mathbf{Y}_{<s}^{(b)}=(y_1^{(b)},\dots,y_{s-1}^{(b)})\) its history. Sampling proceeds as follows:

\emph{(1) Logits adjustment.} Apply penalties~\cite{penalty-fre-pre, penalty-rep} based on \(\mathbf{Y}_{<s}^{(b)}\):
\begin{equation}
    \UseDisplaySpacing
\label{equ:logitsAdj}
\begin{aligned}
\mathbf{Z}'^{(b)}_s
&= \mathrm{ApplyPenalty}\!\bigl(\mathbf{Z}_s^{(b)}, \mathbf{Y}_{<s}^{(b)}\bigr), \ 
\mathbf{Z}'^{(b)}_s \in \mathbb{R}^{V}.
\end{aligned}
\end{equation}

\emph{(2) Probability computation.} Scale by temperature \(\tau\)~\cite{temperature-scaling}, optionally filter candidates using top-\(k\)~\cite{top-k} or nucleus top-\(p_{\mathrm{nuc}}\)~\cite{top-p}, and compute a stable softmax:
\begin{equation}
    \UseDisplaySpacing
\label{equ:probCom}
\begin{aligned}
\tilde{\mathbf{p}}^{(b)}_s
&= \mathrm{softmax}\!\left(
    \mathrm{Filter}\!\bigl(\mathbf{Z}'^{(b)}_s / \tau ;\; k,\, p_{\mathrm{nuc}}  \bigr)
\right).
\end{aligned}
\end{equation}

\emph{(3) Token selection.} Draw the next token ID \(y_s^{(b)} \in \{1,\ldots,V\}\) from the distribution $\tilde{\mathbf{p}}^{(b)}_s$:
\(
y_s^{(b)} \sim \mathrm{Categorical}\!\bigl(\tilde{\mathbf{p}}^{(b)}_s\bigr),
\;
\mathbf{Y}_s \in \mathbb{N}^{B}.
\)

Unlike GEMMs, sampling is dominated by scans over the vocabulary axis: each pass streams $O(V)$ elements with only a handful of operations per element, yielding $O(1)$ FLOPs per byte. Moreover, accesses are \emph{column\mbox{-}major} and often irregular due to masking and index updates, so cache reuse is limited and branch divergence increases. Consequently, sampling saturates memory bandwidth long before compute, making it decisively \emph{memory\mbox{-}bound}.

\subsection{Penalty Algorithm in LLM Sampling}
\label{sec:bg_sample}
We retain the notation from Eq.~\ref{equ:logitsAdj}. Let the per-iteration logits be \(\mathbf{Z}\in\mathbb{R}^{B\times V}\), prompt tokens \(\mathbf{Y}_{\mathrm{p}}\in\mathbb{N}^{B\times L_{\mathrm{p}}}\), and previously generated tokens \(\mathbf{Y}_{<s}\in\mathbb{N}^{B\times (s-1)}\).
We first build per-batch prompt/output histograms: 
\(
\mathbf{C}_{\mathrm{p}}=\mathsf{Hist}(\mathbf{Y}_{\mathrm{p}}),
\mathbf{C}_{\mathrm{o}}=\mathsf{Hist}(\mathbf{Y}_{<s}),
\)
and derive corresponding presence masks \(\mathbf{M}_{\mathrm{p}} = (\mathbf{C}_{\mathrm{p}} > 0)\), \(\mathbf{M}_{\mathrm{o}} = (\mathbf{C}_{\mathrm{o}} > 0)\).
Then construct repetition factors as
\(
\mathbf{f}
= 1+\bigl(\lambda_{\mathrm{rep}}-1\bigr)\,(\mathbf{M}_{\mathrm{p}}\lor \mathbf{M}_{\mathrm{o}}),
\)
where \(\lambda_{\ast}\) are tunable sampling parameters; then
\(\mathbf{Z}' = \mathbf{Z} / \mathbf{f} \).
Other penalties follow analogously and are omitted here for brevity.

\subsection{Vocabulary in LLM Sampling}
\label{sec:bg_vocab}
Modern LLMs adopt markedly larger \(V\) to improve compression on multilingual/code text—e.g., OpenAI models move from \(\sim\!100\mathrm{k}\) (cl100k\_base) to \(\sim\!200\mathrm{k}\) (o200k\_base), Llama~2 uses \(32\mathrm{k}\) while Llama~3 expands to \(128\mathrm{k}\), Gemma reaches \(\sim\!256\mathrm{k}\), and Chinese–English multilingual families such as Qwen (\(\sim\!152\mathrm{k}\)) and Baichuan~2 (\(\sim\!125\mathrm{k}\)) sit in the mid–high range; BLOOM is even larger at \(\sim\!250\mathrm{k}\).
This trend reduces token counts in non-English domains but simultaneously amplifies the memory-bound \(O(V)\) cost of penalties, softmax, and top-\(k\)/\(p_{\mathrm{nuc}}\) scans in \S\ref{sec:bg_sample}, especially under tensor parallelism where global decisions must reconcile shard-local views—motivating designs that avoid full-vocabulary passes~\cite{zhao2025fr, goel2025vocabtrim, zhang2025dynaspec}.

\section{Why Sampling Remains a Holdout}
\label{sec:motivation}

On production traces, we observe that the sampling ratio can reach \textbf{20--38\%} for large vocabularies or constrained decoding, as shown in Figure~\ref{fig:sampling_ratio2}. Increasing \(t\) widens the gap by extra \textbf{10\%}. We also break down per-iteration execution in Figure~\ref{fig:breakdown_iteration_qwen_72B}, showing pipeline bubbles of \(22\%\!-\!40\%\) due to sampling.






Let \(f\!\triangleq\!T_{\text{sampling}}/T_{\text{iter}}\).
If non-sampling work accelerates by a factor \(\rho>1\) (better kernels, parallelism, communication overlap) while the sampling routine is unchanged, then
\begin{equation}
    \UseDisplaySpacing
\label{eq:amdahl}
f'
= \frac{T_{\text{sampling}}}{T_{\text{sampling}} + (T_{\text{iter}}-T_{\text{sampling}})/\rho},
\end{equation}
which increases monotonically with \(\rho\) and satisfies \(f'\!\to\!1\) as \(\rho\!\to\!\infty\). In other words, the sampling fraction \emph{grows} with compute-side speedups (Amdahl’s law).

The reason is structural: mainstream sampling pipelines remain \emph{non-parallelizable} along axes where TP shards. Sampling steps like top-\(k\)/top-\(p_{\mathrm{nuc}}\), penalties, and normalization act along the vocabulary dimension~\footnote{This paper assumes the full set of production sampling controls with fused sampling kernels~\cite{flashinfer_sampling_2025} is enabled.
All these knobs prevent quality confounds as reported in prior work~\cite{top-p, penalty-fre-pre, penalty-rep}.}; with vocab/hidden sharding across \(t\) TP ranks, producing a \emph{global} choice requires consolidating shard-local statistics (e.g., shard top-\(k\) lists or partial CDF prefix sums) via at least one tree reduction, plus an \(O(V/t)\) per-rank scan whose memory-bound cost is insensitive to faster GEMMs. As a result, sampling remains a \emph{serial epilogue}: throughput can scale with more GPUs, yet the per-token latency floor is pinned by \(T_{\text{sampling}}\) and dominates tails (Gustafson’s law).

Moreover, in a \(p\)-stage pipeline, the per-cycle time is:  
\begin{equation}
    \UseDisplaySpacing
\label{eq:cycle}
T_{\text{cycle}}
\;\ge\;
T_{\text{last-compute}} + T_{\text{sampling}},
\end{equation}
because mainstream stacks place sampling at the last stage.

From a Little’s law perspective, two consequences follow. First, throughput is bounded by the pipeline frequency \(1/T_{\text{cycle}}\); adding \(T_{\text{sampling}}\) at the tail caps frequency even when earlier stages are balanced. With deeper pipelines, the aggregate bubble grows because sampling enlarges the stage maximum ($\text{Bubble}=\sum_{i=1}^{p}\bigl(T_{cycle}-T_{\text{stage},i}\bigr)$), thereby increasing idle time across stages. 

\para{Implication.}
Sampling is a durable \emph{holdout}: it neither expands with TP nor balances across PP, and it sets a last-stage latency floor.
A scalable remedy must (1) remove sampling from the GPU critical path, (2) parallelize across sequences (not vocabulary), and (3) overlap its work with the forward pass—precisely the design space \sysname targets.

\section{Design Overview}
\subsection{Challenges and Solutions}
\label{sec:challenges}

\textbf{Challenge: sampling is parallel-unfriendly.}
In tensor parallelism of degree \(t\), parallelizing sampling means sharding the vocabulary dimension
\(\mathcal{V}=\bigsqcup_{r=1}^{t}\mathcal{V}^{(r)}\), each rank \(r\) with shared logits
\(\mathbf{Z}^{(r)}\!\in\!\mathbb{R}^{B\times V/t}\).
Therefore, forming a \emph{global} next-token decision along the vocabulary/hidden axis requires multiple cross-rank reconciliations (and thus collective communication for penalties, softmax, top-\(k\)/\(p_{\mathrm{nuc}}\)), causing extra overhead of several milliseconds. 
Meanwhile, penalties depend on per-sequence histories \((\mathbf{C}_{\mathrm{o}},\mathbf{M}_{\mathrm{o}})\), creating fine-grained data dependencies and metadata movement.
Under pipeline parallelism of depth \(p\), mainstream stacks execute sampling only at the last stage, elongating \(T_{\text{stage},p}\) and capping the pipeline frequency.
Dedicating a GPU to sampling is wasteful (low arithmetic intensity, bandwidth-bound) and couples model partitioning with decision logic; conversely, a \emph{naïve} CPU offload often fails to fit within \(T_{\text{stage},p}\) without overlap-aware design.

\textbf{Solution 1: \emph{Sequence-parallel sampling}.}
\sysname scales the decision plane under TP by partitioning batch indices \( \{1,\ldots,B\}\) into disjoint \(B/m\) blocks, assigning to $m$ samplers.
TP ranks continue producing sharded logits \(\mathbf{Z}_{:,1\!:\!B}\in\mathbb{R}^{B\times V/t}\) into the shared memory so that samplers never perform copies to derive their blocks.
Per-sequence metadata \((\mathbf{C}_{\mathrm{o}},\mathbf{M}_{\mathrm{o}})\) follow the same batch partition and are
updated locally.
This avoids global \(V\)-axis decisions and turns sampling into
independent per-sequence tasks that scale with the number of workers.

\textbf{Solution 2: \emph{Column-wise penalties with truncation-first filtering on CPUs}.}
To remove the last-stage skew under PP, \sysname offloads the decision plane to CPUs but avoids a naïve \(O(V)\) full pass via:
(i) a column-wise layout that supports in-place, incremental updates of the penalty states without rebuilding large tensors; and
(ii) a \emph{truncation-first} pass (top-\(k\)/\(p_{\mathrm{nuc}}\)) that narrows the active set before normalization.
These choices cut memory traffic and enable overlap with GPU compute.

\textbf{Solution 3: \emph{Speculative hot-vocab sampling with rejection-correctness}.}
For large \(V\), CPU-offloaded sampling can still dominate. Motivated by Zipf-like token distributions, \sysname
constructs a model-dependent hot set \(\mathcal{H}\subset\mathcal{V}\) (size \(H=|\mathcal{H}|\ll V\)).
For each sequence \(b\), we sample on \(\mathcal{H}\) (fast path) and apply \emph{rejection sampling} against the full
\(\tilde{\mathbf{p}}\) to preserve exactness. Let the covered mass be
\(\alpha_b=\sum_{v\in \mathcal{H}}\tilde{p}_{b,v}\).
The approach yields high acceptance (80–95\% in our traces)
and substantial decision-plane speedups while keeping the output distribution unbiased.

\subsection{System Architecture and Workflow}
\label{sec:overview}

\begin{figure}[t]
    \centering
    \UseDisplaySpacing
    \includegraphics[width=0.95\linewidth]{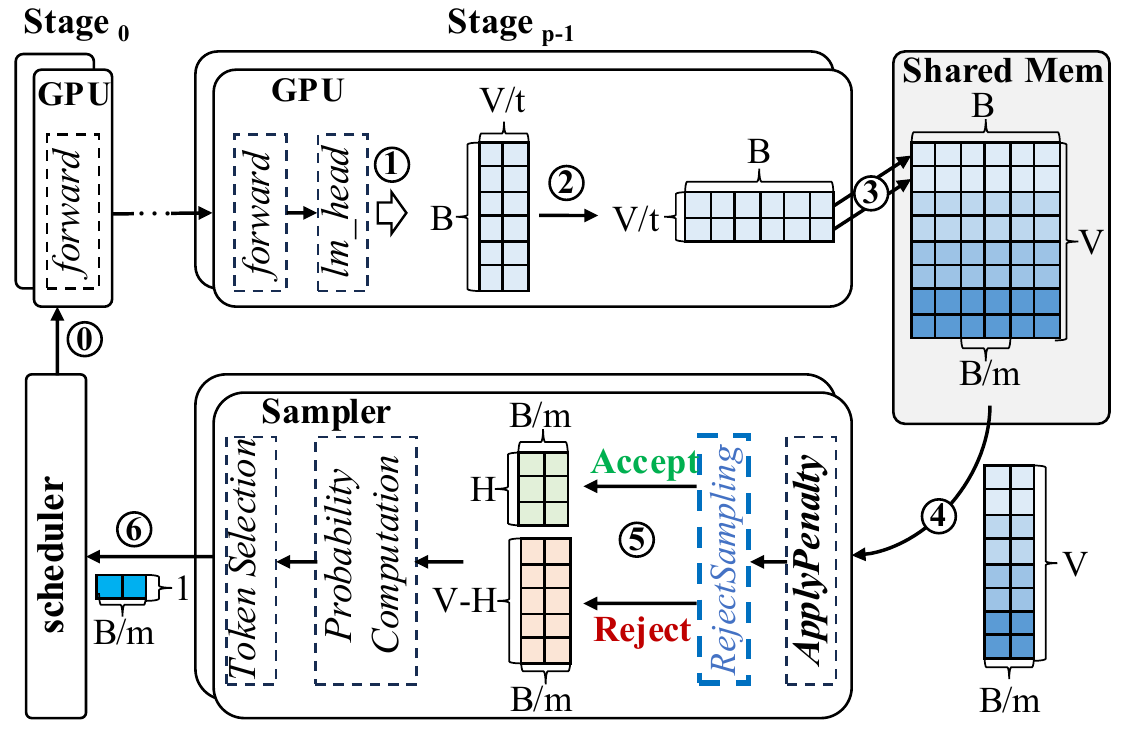}
    \caption{Architecture and workflow of \sysname.}
    \label{fig:overview}
\end{figure}

\sysname disaggregates sampling from the main inference engine as an independent CPU-side service. The system comprises three roles—\emph{scheduler}, \emph{GPU workers}, and \emph{CPU samplers} (as in Figure~\ref{fig:overview}).

\textbf{Data flow.}
Shared-memory ring buffers carry three streams:
(i) compact \emph{scheduling outputs} (sequence IDs, KV pointers, iteration \(s\), masks),
(ii) TP-sharded \emph{logits blocks} written by the final-stage GPU workers, and
(iii) auxiliary sampler inputs (e.g., pre-generated random numbers).
Producers and consumers advance independently for better overlap.

In multi-host deployments, scheduling outputs are sent to a worker on downstream hosts over NCCL first and then copied to the shared memory. 
\sysname assumes that, for each pipeline, the \emph{last PP stage} and all of its TP ranks are co-located on the same host, so logits are never reconstructed across the network.
Samplers write back decisions (next-token IDs, EOS flags, optional log-probs) via a lightweight ZMQ~\cite{hintjens2013zeromq} channel to the scheduler.

\textbf{End-to-end workflow.} The per-iteration workflow of \sysname can be seen as follows:

\emph{\num{0} Scheduling.} The scheduler selects the active microbatch of size \(B\), emits one \emph{scheduling output}, and enqueues it to all GPU workers and CPU samplers.

\emph{\num{1} GPU compute.} Each GPU worker executes its model partition. Non-final PP stages forward activations of shape \(B\times h\). Final-stage workers produce TP-sharded logits for the current microbatch,
\(\mathbf{Z}_{:,1\!:\!B}\in\mathbb{R}^{B\times V/t}\). 

\emph{\num{2} Transpose to vocabulary-major.} Final-stage workers transpose sharded logits to \([V/t \times B]\) to favor column-wise scans and contiguous writes for downstream sampling.

\emph{\num{3} Logits write.} Each final-stage worker writes its rank-local block \([V/t \times B]\) into shared memory. Logically, the microbatch forms a global \(V\times B\) matrix composed of \(\{V/t\}\) vertical slices.

\emph{\num{4} Sequence-parallel read.} With \(m\) CPU samplers, partition the \(B\) columns into disjoint index sets \(B_1,\dots,B_m\) with \(|B_j|\approx B/m\). Sampler \(j\) reads the block \([V/t \times |B_j|]\) for every TP rank  and concatenates along the vocabulary axis to reconstruct \(\mathbf{Z}_{:,B_j}\in\mathbb{R}^{V\times |B_j|}\) in a zero-copy manner, because sampling operates directly on shared memory. Per-sequence metadata (histograms/masks) for \(B_j\) are read from the metadata ring.

\emph{\num{5} SHVS decision.} For each sequence \(b\in B_j\), the sampler applies penalties and performs \emph{speculative hot-vocab sampling} with \emph{rejection-correctness} to select \emph{which sub-vocabulary} (hot or tail) proceeds to full decision.

\emph{\num{6} Commit.} Samplers write decisions to the return ring. The scheduler commits results to engine state, retires finished sequences, and issues the next microbatch.

\section{Design}
\label{sec:design}

This section turns \sysnames principles—\emph{parallelizable}, \emph{stage-agnostic}, and \emph{overlappable} sampling—into three concrete mechanisms that remove sampling from the GPU critical path while preserving \emph{distribution-exact} semantics.

\subsection{Sequence-Parallel Sampling}
\label{sec:sampling}

While TP accelerates the \emph{forward} data plane, production stacks still execute sampling as a single-node epilogue. Yet operations on \(\mathbf{Z}\!\in\!\mathbb{R}^{B\times V}\) are independent \emph{across sequences} (the batch/row axis): each row depends only on its own history and masks. \sysname exploits this by sharding the sampling workload over \(m\) samplers along the batch indices, forming local decisions \emph{without} vocabulary-axis collectives.

\textbf{Workflow.}
Per iteration, the scheduler broadcasts a compact \emph{scheduling output} to all workers and samplers. Final-stage GPU workers write TP-sharded logits in a vocabulary-major layout \([V/t\times B]\) into shared-memory rings; \(m\) CPU samplers then partition the \(B\) rows into disjoint sets \(B_1,\dots,B_m\) with \(|B_j|\!\approx\!B/m\) and, for each \(B_j\), reconstruct full-vocabulary views by concatenating the rank-local slices \([V/t\times |B_j|]\) across \(r=1,\dots,t\). Because samplers operate directly on the shared-memory buffers, this reconstruction is realized via tensor views/strides without copies. 
Each sampler completes penalties, filtering, and token draws for its \(B_j\) independently and returns decisions to the scheduler. This turns sampling into \(m\) independent per-sequence tasks and avoids any \(\mathrm{AllGather}_V\) on the critical path.

\textbf{Deterministic random number generation.}
Sampling \emph{uses} random variates for the final categorical draw.
Naïvely parallel random number generators (RNGs) can diverge from single-worker outcomes. To retain determinism under a fixed seed, \sysname pre-generates random numbers on all $t$ GPUs and lets each sampler consume its slice via shared memory as they do on logits. This approach adds negligible overhead: samplers read their random-number slices asynchronously via shared memory, and retrieving one slice takes <1 ms in our measurements.

\subsection{Advanced CPU Sampling Algorithm}

\label{sec:cpu}
Sequence parallelism scales sampling under TP, but with PP, the last stage still elongates if sampling remains on GPUs. Offloading the decision plane to CPUs decouples it from the pipeline; however, a naïve port cannot be fully overlapped with the \emph{forward} pass (see \S\ref{sec:microbench}). \sysname therefore adopts CPU-centric algorithms to accelerate sampling. Our evaluation also proves that \sysname introduces little CPU overhead even after offloading (\S\ref{sec:util}). 

\textbf{Vocabulary-major, column-wise layout for penalties.}
We store per-iter tensors in a \emph{vocabulary-major} (column-wise) form by transposing \(\mathbf{Z}\in\mathbb{R}^{B\times V}\) and \(\mathbf{Y}_{<s}\in\mathbb{N}^{B\times (s-1)}\) to \(\mathbf{Z}^{\top}\in\mathbb{R}^{V\times B}\) and \(\mathbf{Y}^{\top}\in\mathbb{N}^{(s-1)\times B}\). This layout is reused across steps to build penalties incrementally and to streamline downstream filtering, improving both compute efficiency and memory locality.

We preallocate a row-append buffer \(\mathbf{Y}\in\mathbb{N}^{L_{\max}\times B}\) for generated tokens and write the step-\(s\) output as row \(s\).
Contiguous row appends avoid tensor reconstruction and allocator churn.
Instead of rebuilding histograms, we update counts using only the new row.
Let \(\mathbf{C}_{\mathrm{o}}\in\mathbb{N}^{B\times V}\) be the output histogram; then
\begin{equation}
    \UseDisplaySpacing
\label{eq:cpu-incremental}
\begin{aligned}
\mathbf{C}_{\mathrm{o}}^{\,s+1} &= \mathbf{C}_{\mathrm{o}}^{\,s} + \mathsf{Hist}(\mathbf{Y}_s),\\
\mathbf{M}_{\mathrm{o}}^{\,s+1} &= \bigl(\mathbf{C}_{\mathrm{o}}^{\,s+1} > 0\bigr),
\end{aligned}
\end{equation}
and the repetition factor \(\mathbf{f}\) follows \S\ref{sec:bg_sample} (with \(\mathbf{C}_{\mathrm{p}}\) step-invariant).
Because only the newest row is touched, updates are cache-friendly.

\textbf{Truncation-first filtering.}
Beyond penalty computation, softmax can also become a CPU\-side bottleneck. 
Let the filter compose all enabled constraints (top-$k$, nucleus, allow list, etc.) into a per-sequence subset
\(\mathcal{K}_b = \mathrm{Filter}(\mathcal{V};\,k,\,p_{\mathrm{nuc}},\ldots)\).
We first build an \emph{index map} from the full vocabulary to this subset: $\pi_b:\{1,\ldots,|\mathcal{K}_b|\}\!\to\!\{1,\ldots,V\}$,
so \(\pi_b\) maps subset indices back to \(\mathcal{V}\) and \(\pi_b^{-1}\) provides a fast \(V\!\to\!\mathcal{K}_b\) lookup.
We then \emph{truncate} logits to the selected domain and normalize only there: $\tilde{\mathbf{p}}^{(b)}_s = \mathrm{softmax}\!\bigl( (\mathbf{Z}'^{(b)}_s/\tau)\big|_{\mathcal{K}_b} \bigr)$
and finally map the sampled index back to the full vocabulary by \(\tilde{y_s}^{(b)}=\pi_b(y_s^{(b)})\).
This implementation preserves \emph{exact} semantics: softmax on \(\mathcal{K}_b\) equals masked softmax over \(\mathcal{V}\) but reduces complexity from $O(V)$ to $O(k)$, cutting CPU work and memory traffic. 

\subsection{Speculative Hot-vocab Sampling}
\label{sec:hotvocab}

Even with sequence parallelism and advanced algorithms, the \(O(V)\) scans in top-\(k\)/top-\(p_{\mathrm{nuc}}\)/penalty applying can dominate at modern vocabulary sizes ($\ge$ 10K).
To keep sampling off the critical path, \sysname must keep per-iter sampling time below the longest GPU stage—and ideally below on-GPU sampling time—to avoid degrading pipeline frequency or per-token latency.

Empirically, next-token probabilities are \emph{Zipf-like}: a small head of tokens carries most probability mass (e.g., top \(32\mathrm{k}\) often covers \(>\!95\%\)) across different workloads, which means
we can build a model-dependent  hot set (i.e., a small sub-vocabulary) \(\mathcal{H}\subset\mathcal{V}\) with \(H\triangleq|\mathcal{H}|\ll V\) using offline traces~\cite{zhao2025fr, goel2025vocabtrim, zhang2025dynaspec}.
\sysname exploits this via \emph{speculative hot-vocab sampling (SHVS)}.
Let \(\mathbf{Z}'_s\in\mathbb{R}^{B\times V}\) be the penalized logits at step \(s\).
For sequence \(b\in\{1,\dots,B\}\) and token \(v\in\mathcal{V}\), define numerically stable weights
\begin{equation}
    \UseDisplaySpacing
\label{eq:hot-weights-z}
\begin{aligned}
w_{b,v}
&= \exp\!\Bigl(Z'^{(b)}_{s,v} - \max\nolimits_{i\in\mathcal{V}} Z'^{(b)}_{s,i}\Bigr).
\end{aligned}
\end{equation}
$w_{b,v}$ can be pre-computed on GPUs when writing logits. 
Split the support into the hot set \(\mathcal{H}\) and tail \(\mathcal{V}\setminus\mathcal{H}\), and compute the \emph{hot-vocab mass} ($\alpha_b$):
\begin{equation}
    \UseDisplaySpacing
\label{eq:hot-masses-z}
\begin{aligned}
S_{b,\mathcal{H}} &= \sum_{v\in \mathcal{H}} w_{b,v}, \quad
S_{b,\mathcal{V}\setminus\mathcal{H}} = \sum_{v\in \mathcal{V}\setminus\mathcal{H}} w_{b,v}, \\
\alpha_b &= \frac{S_{b,\mathcal{H}}}{S_{b,\mathcal{H}}+S_{b,\mathcal{V}\setminus\mathcal{H}}}, \quad
\tilde{p}_{b,v} = \frac{w_{b,v}}{S_{b,\mathcal{H}}+S_{b,\mathcal{V}\setminus\mathcal{H}}}.
\end{aligned}
\end{equation}

\para{Rejection sampling.}
Define hot and tail proposals
\begin{equation}
    \UseDisplaySpacing
\label{eq:hot-q-r}
\begin{aligned}
q_{b,v} = \frac{w_{b,v}}{S_{b,\mathcal{H}}}, v\in \mathcal{H}; \quad
r_{b,v} = \frac{w_{b,v}}{S_{b,\mathcal{V}\setminus\mathcal{H}}}, v\in \mathcal{V}\setminus\mathcal{H}.
\end{aligned}
\end{equation}
Draw a hot candidate \(\hat{y}\sim q_{b,\cdot}\) and a uniform variate \(u\sim\mathrm{Unif}(0,1)\).
Accept \(\hat{y}\) if \(u\le \alpha_b\); otherwise \emph{reject} and draw \(y'\sim r_{b,\cdot}\).
Return \(y_b=\hat{y}\) on acceptance and \(y_b=y'\) on rejection.
Since \(\tilde{p}_{b,v}/q_{b,v}=\alpha_b\) for all \(v\in\mathcal{H}\), this is rejection sampling with envelope \(M{=}1\) on the hot path, and
\begin{equation}
    \UseDisplaySpacing
\label{eq:hot-correctness}
\begin{aligned}
\mathbb{P}[y_b = v]
&= \alpha_b\, q_{b,v}\,\mathbf{1}[v\in \mathcal{H}] \\
&+ (1-\alpha_b)\, r_{b,v}\,\mathbf{1}[v\in \mathcal{V}\setminus\mathcal{H}]
\\
&= \tilde{p}_{b,v},
\end{aligned}
\end{equation}
i.e., \emph{distributionally exact}.

\para{Determinism and mapping.} We use a fixed-seed RNG so that \(u\) is reproducible across samplers. The output of SHVS is a double-indexing result on the filtered probabilities of the sub-vocabulary, so SHVS remaps results back to the full hot-vocab first using the top-$k$ indices and then to the full vocabulary using the hot/tail indices.

\subsection{Choosing the Hot-vocab Size}
\label{sec:hotset-sizing}

Finally, we discuss how to select $H$ for SHVS. The hot vocab size \(H\) trades off two opposing effects: (i) a larger \(H\) increases hot-path work, but (ii) raises the covered mass \(\alpha_b\) so fewer sequences fall back to the full-vocabulary tail. We choose \(H\) to maximize end-to-end throughput using an offline, hardware-aware cost model composed with an empirical hit-rate model. Note that throughput tuning does not affect distributional exactness because \S\ref{sec:hotvocab} uses rejection-correctness irrespective of \(H\).

\textbf{Hit-ratio model.} Define the hot-vocab mass per sequence \(b\) as \( \alpha_b(H) \triangleq \sum_{v\in H}\tilde{p}_{b,v} \) and the average hit ratio $\bar{\alpha}(H) \triangleq \mathbb{E}_b\big[\alpha_b(H)\big]$.
Empirically, \(\bar{\alpha}(H)\) is monotone and Zipf-like—dominated by the model (and decoding policy) rather than hardware—so it can be modeled offline from traces.

\textbf{CPU-sampling cost model.}
SHVS deliberately uses \emph{single-pass} scans, so time grows \emph{linearly} with the number of tokens visited in vocabularies.
The expected sampling time per sequence on $\mathcal{H}$ is $T_{\mathrm{cpu}}(H) = cH + c_0$,
leading to an \emph{affine} CPU-time model
\begin{equation}
    \UseDisplaySpacing
\label{eq:cpu-time}
\begin{aligned}
F(H) = \mathbb{E}\!\left[T_{\mathrm{cpu}}(H)\right]
&\approx c_0 + c\,
\Big(
\bar{\alpha}(H)\,H
\\[-1pt]
&+ \bigl(1-\bar{\alpha}(H)\bigr)\,(V-H)
\Big).
\end{aligned}
\end{equation}

The constants \(c_0\) and \(c\) are platform-specific; a few points suffice to fit them (see \S\ref{sec:sizing-validate}).

\textbf{Optimal \(H\) via first-order condition.}
Differentiating Eq.~\eqref{eq:cpu-time} w.r.t.\ \(H\) gives
\begin{equation}
    \UseDisplaySpacing
\label{eq:first-derivative-final}
\begin{aligned}
\frac{dF}{dH}
&=\;
c\Big(
 -1
 + 2\,\bar{\alpha}(H)
 + (2H - V)\,\bar{\alpha}'(H)
\Big).
\end{aligned}
\end{equation}
Setting the derivative to zero yields the stationary point \(H^\star\):
\begin{equation}
    \UseDisplaySpacing
\label{eq:stationary-h}
\begin{aligned}
2\,\bar{\alpha}(H^\star)
\;+\;
(2H^\star - V)\,\bar{\alpha}'(H^\star)
\;=\;
1.
\end{aligned}
\end{equation}
\textit{Interpretation.} Increase the hot-vocab size until the marginal reduction of expected tail work
\(\,(2H{-}V)\bar{\alpha}'(H) + 2\bar{\alpha}(H)\) balances the unit cost of growing \(H\) (the RHS).
Under the monotone, saturating (empirically concave) \(\bar{\alpha}\), this condition admits a unique interior solution.
Choosing \(H\approx H^\star\) minimizes CPU sampling time; because \(H\) is discrete (\(H\in\{1,\ldots,V\}\)), we enumerate around the continuous optimum and choose $arg\,min_HF(H)$ in deployment. 
\begin{table}[t]
\caption{Evaluation testbed overview.}
\label{tab:testbed}
\centering
\renewcommand{\arraystretch}{1.12}
\setlength{\tabcolsep}{3pt}
\begin{tabularx}{\columnwidth}{|>{\raggedright\arraybackslash}p{0.26\columnwidth}|X|X|X|}
\hline
\textbf{Config} & \textbf{L40 node} & \textbf{H100 node} & \textbf{B200 node} \\
\hline
GPU model & NVIDIA L40 & NVIDIA H100 & NVIDIA B200 \\
\hline
GPU memory & 48 GB & 80 GB & 180 GB \\
\hline
Intra-node interconnect & PCIe 4.0 & NVLink & NVLink \\
\hline
Inter-node network & 200 Gbps & $8\times 400$ Gbps & $8\times 400$ Gbps \\
\hline
CPU & 128$\times$ Intel Xeon Platinum 8358 & 192$\times$ Intel Xeon Platinum 8468 & 256$\times$ Intel Xeon 6767P \\
\hline
CUDA version & 12.6 & 12.6 & 12.8 \\
\hline
\end{tabularx}
\end{table}

\newcolumntype{Y}{>{\raggedright\arraybackslash}X}
\newcommand{\NA}{\textemdash}              
\newcommand{\tppt}[2]{TP#1--PP#2}           

\begin{table}[t]
\centering
\caption{Models and TP/PP degrees per platform. (\NA) indicates the model is not evaluated on that platform: (i) it is too large and requires $>$16 GPUs, or (ii) it is small enough to run on a single GPU without distribution.}
\label{tab:model-topology}
\begin{threeparttable}
\begin{tabularx}{\linewidth}{Y c c c}
\toprule
\textbf{Model} & \textbf{L40} & \textbf{H100} & \textbf{B200} \\
\midrule
QwQ-32B                   & \tppt{4}{1} & \NA         & \NA         \\
Llama-3.1-70B             & \tppt{4}{2} & \tppt{4}{2} & \NA         \\
Qwen-2.5-72B              & \tppt{4}{2} & \tppt{4}{2} & \NA         \\
Qwen3-235B-A22B           & \tppt{4}{4} & \tppt{4}{4} & \tppt{4}{2} \\
DeepSeek V3               & \NA         & \tppt{4}{4} & \tppt{4}{2} \\
Qwen3-Coder-480B-A35B     & \NA         & \NA         & \tppt{4}{2} \\
\bottomrule
\end{tabularx}
\end{threeparttable}
\end{table}

\section{Implementation}
\sysname is implemented in Python and integrates as a drop-in extension to \texttt{vLLM}~\cite{kwon2023vllm}. It subclasses the engine’s scheduling and sampling interfaces and overrides only the decision-plane hooks, requiring \emph{no} changes to the upstream \texttt{vLLM} source. The codebase is lightweight (\(\sim\)6K lines of Python).
At runtime, the plugin launches a co-scheduled CPU sampler group and a set of shared-memory ring buffers for zero-copy exchange of logits slices and sampling decisions. 
The implementation preserves full sampling functionality (temperature, top-\(k\), nucleus top-\(p_{\mathrm{nuc}}\), repetition/presence/frequency penalties, and optional biases). SHVS loads a hot sub-vocabulary \(\mathcal{H}\) at startup and exposes a lightweight control to adjust \(H\) using the sizing rule from \S\ref{sec:hotset-sizing}. Basic observability (acceptance rate \(\alpha\), sampler throughput, and overlap statistics) is included to aid tuning in production.

\begin{figure*}[!t]
  \centering
  \includegraphics[width=\textwidth]{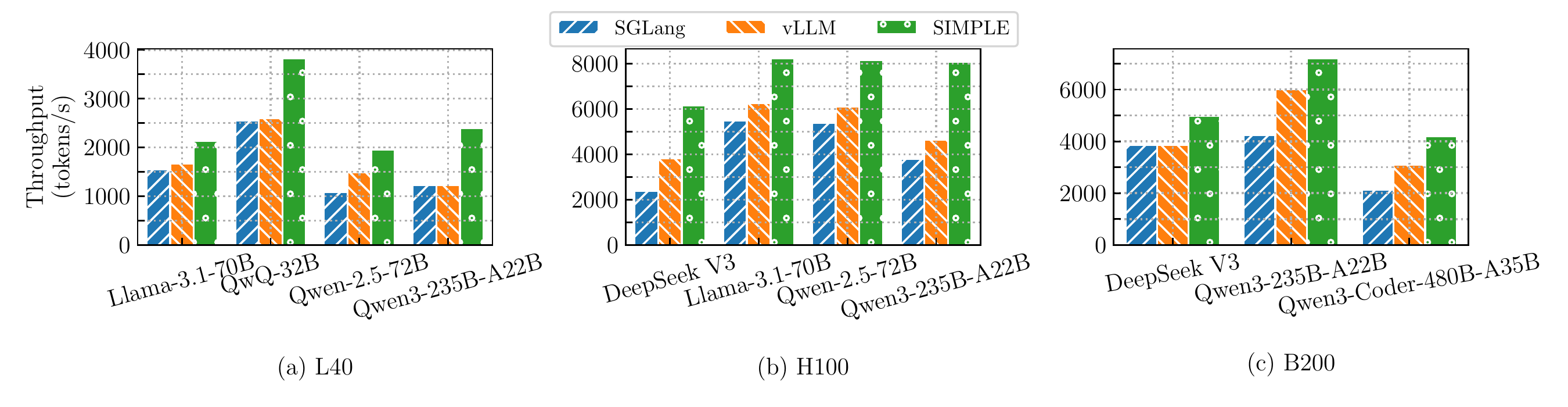}
  \caption{End-to-end throughput (tokens/s) across platforms and models.}
  \label{fig:throughput_all}
\end{figure*}

\section{Evaluation}

This section evaluates \sysname across models and testbeds and answers four questions:
(i) Does it improve end-to-end throughput and per-token latency?
(ii) How do GPU/CPU utilizations change under our design?
(iii) Which decision-plane components dominate CPU time?
(iv) Does SHVS remain exact, and does the sizing model in \S\ref{sec:hotset-sizing} predict the optimal \(H^\star\)?

\subsection{Evaluation Setup}
\label{sec:exp_setup}

\textbf{Testbed.} Our evaluation is conducted on the following testbeds: {\tt L40}, {\tt H100}, and {\tt B200}. Each server is equipped with 8 GPUs and 2\,TB of host memory. Detailed hardware specs are provided in Table~\ref{tab:testbed}.

\textbf{Baseline.} We compare \sysname against two widely used inference engines: {\tt vLLM} (0.10.1)~\cite{kwon2023vllm} and {\tt SGLang} (0.5.2)~\cite{zheng2024sglang}. 

\textbf{Models and deployments.} We select models that are served in distributed configurations on each platform. Table~\ref{tab:model-topology} lists the TP/PP degrees that maximized throughput under our constraints.
Following common practitioner guidance~\cite{vllm_tunning}, we cap tensor parallelism at \(t\!\le\!4\) to preserve scaling efficiency across nodes. On the decision plane, we use 16 samplers for each engine and 4 threads for each sampler. 

\textbf{Workload.}
To ensure fair and reproducible comparisons, we replay a fixed prompt set sampled from ShareGPT~\cite{shareGPT} and disable early stopping.
We enable the full production sampling controls—top-\(p_{\mathrm{nuc}}\), top-\(k\), min-\(p\), temperature, and repetition/presence/frequency penalties—to avoid quality confounds.
Unless stated otherwise, the default per-GPU batch size is \(B{=}32\) (e.g., total batch size = 256 when $p \times t = 8$).

\begin{figure}[t]
  \centering
  \includegraphics[width=0.95\columnwidth]{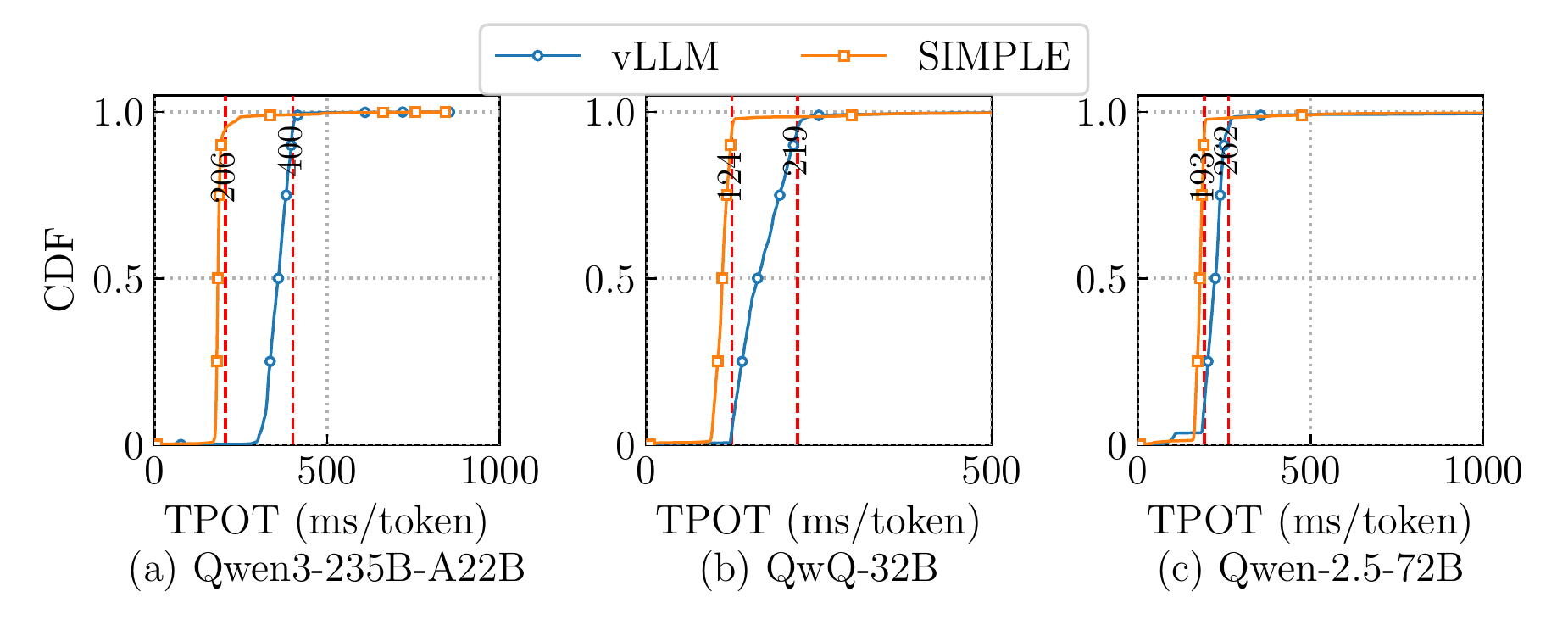}
  \caption{TPOT ECDF on L40 (P95 marked).}
  \label{fig:tpot_l40}
\end{figure}

\subsection{End-to-end Scaling and Performance}
\label{sec:e2e}

\textbf{Throughput across models and testbeds.}
Across devices, \sysname delivers substantial end-to-end throughput gains over on-GPU baselines (Figure~\ref{fig:throughput_all}).
On L40, the average improvement over \texttt{vLLM} is \textbf{+50\%}, peaking at \textbf{+96\%} on \textit{Qwen3-235B-A22B}.
On H100, the average gain is \textbf{+50\%} with a maximum of \textbf{+74\%} on \textit{Qwen3-235B-A22B}.
On B200, improvements are smaller but still material (mean \textbf{+28\%}; max \textbf{+36\%} on \textit{Qwen3-Coder-480B-A35B}).
Against another optimized stack (\texttt{SGLang}), \sysname also leads by large margins (e.g., on H100: \textbf{+67\%} with \textit{Llama-3.1-70B}), indicating that benefits are not tied to a specific baseline.

These results align with our design hypotheses: SHVS cuts the common-case decision work from \(O(V)\) to \(O(H)\) with \(H\!\ll\!V\), so models with larger vocabularies (e.g., \textit{Qwen3-235B}, \textit{480B}) see the largest gains, matching the Zipf-driven cost model in \S\ref{sec:hotvocab}–\S\ref{sec:hotset-sizing}. We also notice that \textit{Qwen3-235B-A22B} gain on L40 is especially large for two reasons: (i) deeper pipelining (\(p{=}4\)) inflates the last-stage skew in the baseline: a lager \(T_{\text{cycle}}\) accumulates latency along stages. (ii) in multi-host deployments \sysname avoids cross-machine broadcast on \emph{scheduling output} but fans out \emph{intra-host} through shared-memory rings (\S\ref{sec:overview}).  

Higher throughput allows operators to meet a given load with fewer GPUs or to tighten latency SLOs at similar cost.
The architecture composes with data-plane optimizations and continues to amortize sampling as accelerators improve.

\begin{figure}[t]
  \centering
  \includegraphics[width=\columnwidth]{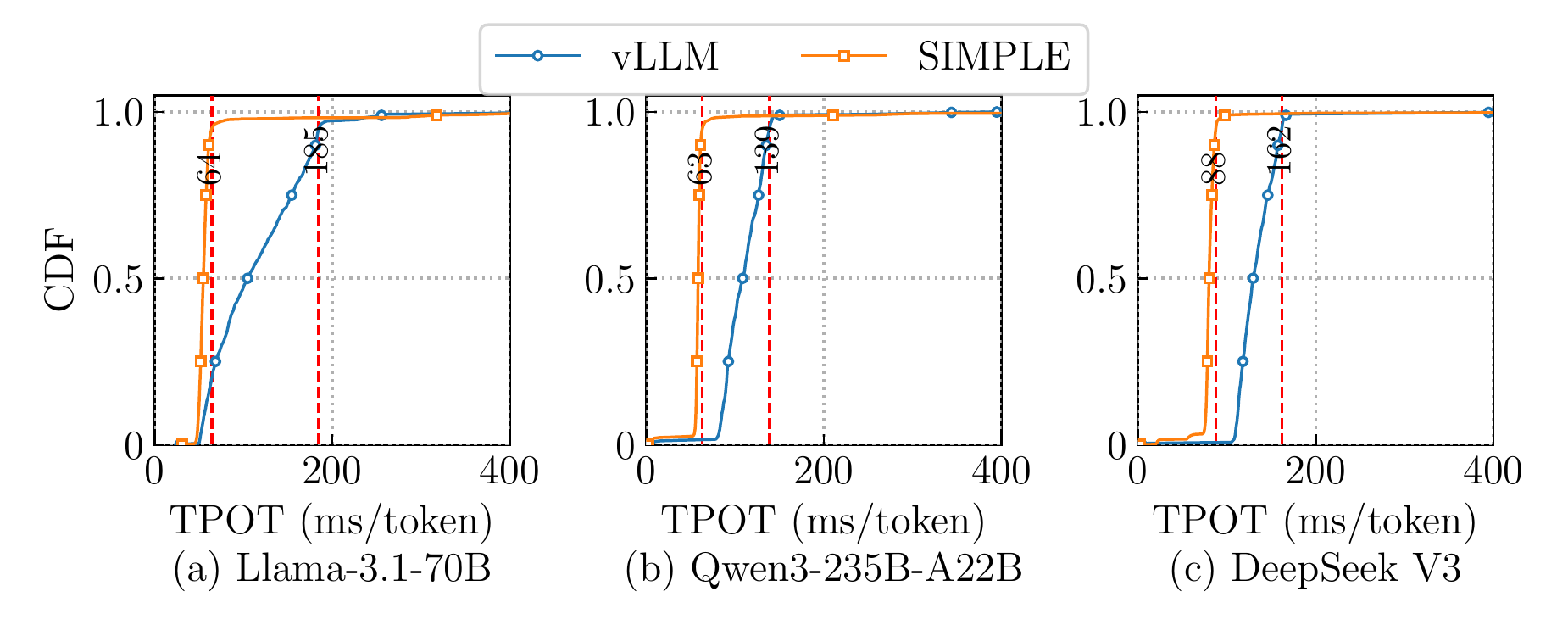}
  \caption{TPOT ECDF on H100 (P95 marked).}
  \label{fig:tpot_h100}
\end{figure}

\begin{figure}[t]
  \centering
  \includegraphics[width=\columnwidth]{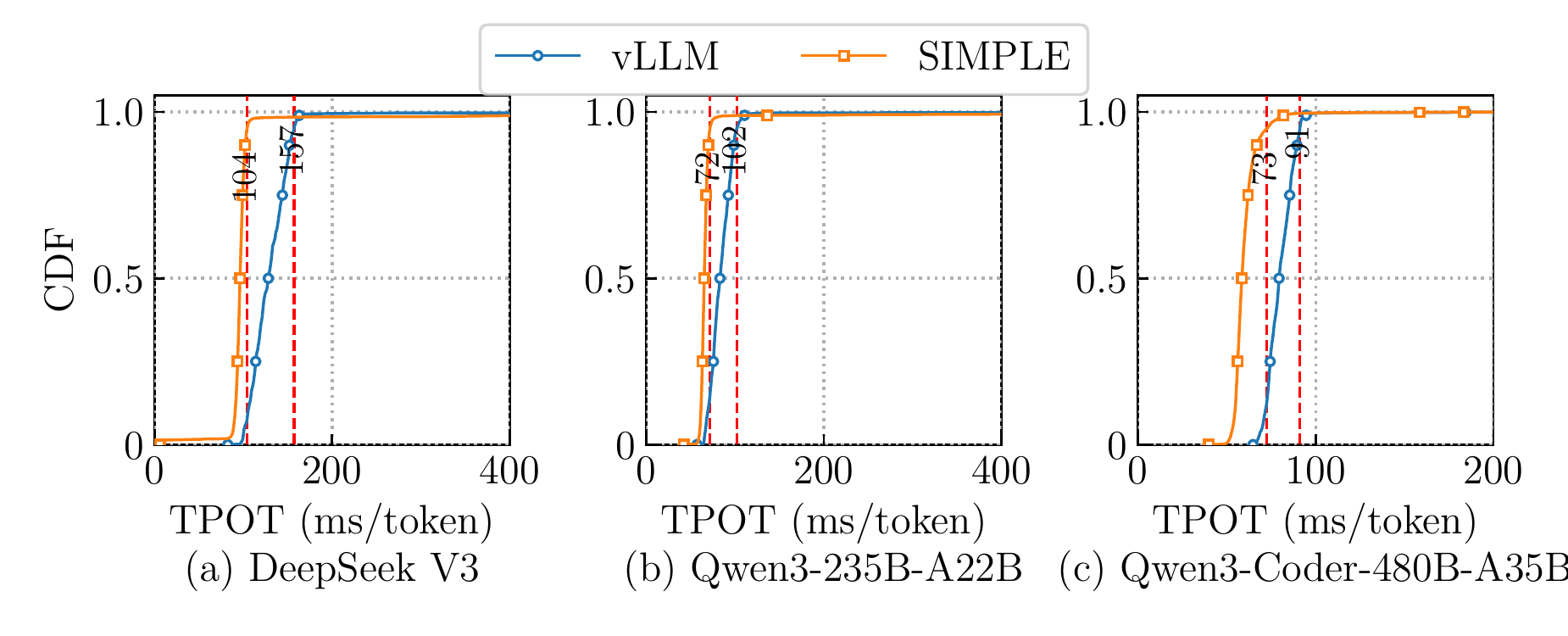}
  \caption{TPOT ECDF on B200 (P95 marked).}
  \label{fig:tpot_b200}
\end{figure}

\textbf{Latency across models and testbeds.}
Time-to-First-Token (TTFT) is primarily governed by scheduling/queuing and is unaffected by \sysname.
We therefore focus on Time-per-Output-Token (TPOT) at P95 under default configurations.
\sysname consistently lowers tails across devices:
on H100, mean P95 reduction is \textbf{55\%} (max \textbf{65\%} on \textit{Llama-3.1-70B}; see Figure~\ref{fig:tpot_h100});
on L40, mean \textbf{39\%} (max \textbf{49\%} on \textit{Qwen3-235B-A22B}; Figure~\ref{fig:tpot_l40});
on B200, mean \textbf{28\%} (max \textbf{34\%} on \textit{DeepSeek~V3}; Figure~\ref{fig:tpot_b200}).

These tail gains validate the design claims: by making the decision plane \emph{stage-agnostic} and \emph{overlappable}, \sysname removes the last-stage epilogue from the pipeline’s critical path and hides decision latency beneath GPU compute.
As a result, P95 shrinks even as \(t\) and \(p\) scale and as GPUs get faster—counteracting the Amdahl drift that inflates the baseline’s sampling fraction \(f\) (Eq.~\ref{eq:amdahl}) and enabling tighter SLOs or higher admission rates at the same budget.

\begin{figure}[tb]
  \centering
  \includegraphics[width=\columnwidth]{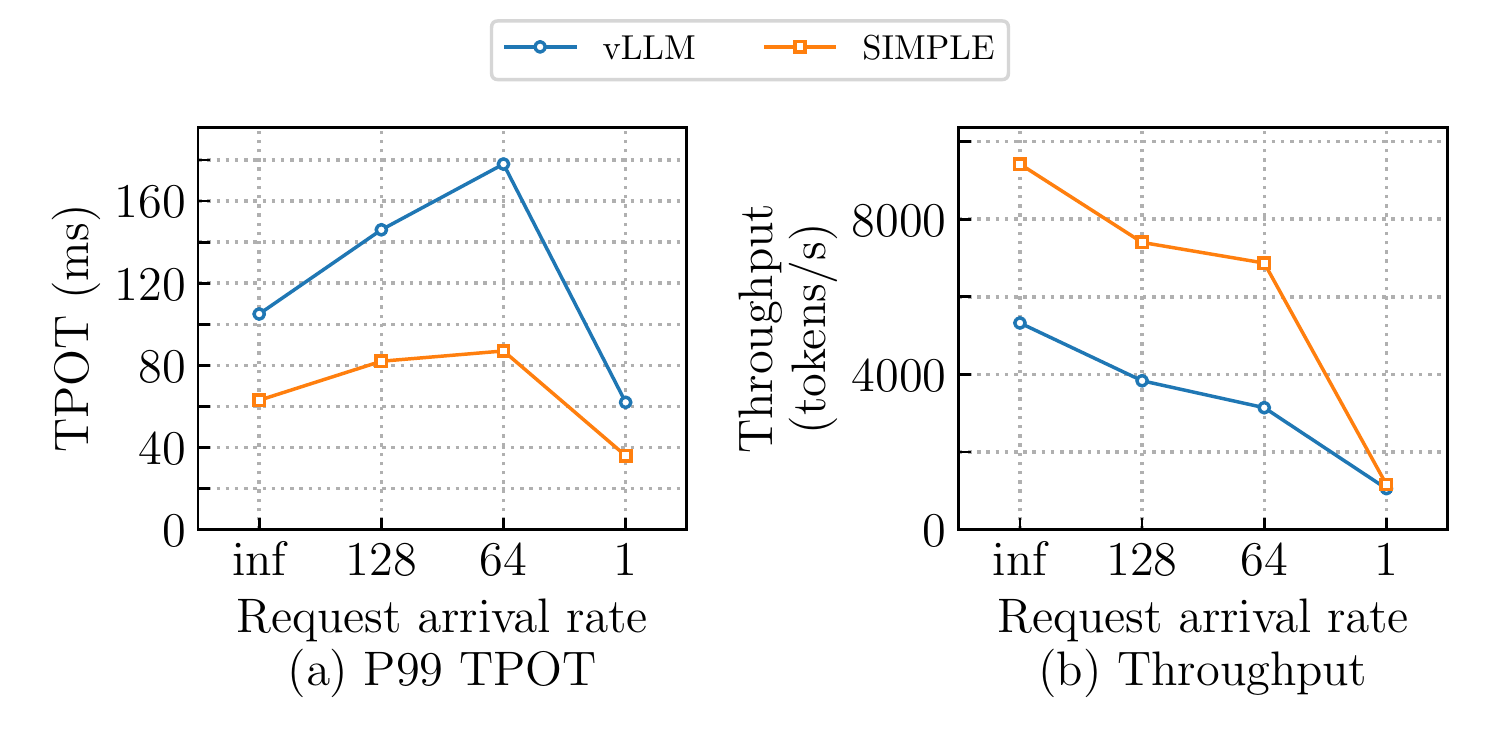}
  \caption{TPOT/throughput v.s. request rate on H100 with Qwen3-235B-A22B.}
  \label{fig:tpot_th}
\end{figure}

\para{Load–latency tradeoff.}
Varying the request arrival rate reveals a consistent right-and-up shift of the throughput–P99 curve under \sysname relative to vLLM on H100 with \textit{Qwen3-235B-A22B} in Figure~\ref{fig:tpot_th}. 
At saturation (\texttt{rate=inf}), \sysname cuts P99 TPOT from \textbf{105}\,ms to \textbf{63}\,ms (\textbf{-40\%}) while boosting throughput from \textbf{5326} to \textbf{9421} tok/s (\textbf{+77\%}, \(\sim\!1.8\times\)). 
At moderate loads, \texttt{rate=64} achieves \textbf{-51\%} P99 (178\(\rightarrow\)87\,ms) and \textbf{+119\%} throughput (3143\(\rightarrow\)6870 tokens/s, \(\sim\!2.2\times\)). 
Even at very low load (\texttt{rate=1}), where batching effects dominate, \sysname halves the tail (62\(\rightarrow\)36\,ms, \textbf{-42\%}), indicating that the GPU cadence is no longer gated by a serial epilogue.

For a fixed P99 SLO, \sysname admits substantially higher arrival rates. 
At saturation, the \(\sim\!1.8\times\) throughput gain means fewer GPUs for the same capacity or lower queueing delay at the same budget. 
At low load, eliminating last-stage bubbles maintains high GPU residency even without large batches, improving tail latency for interactive workloads.

The baseline’s P99 is non-monotonic between \texttt{rate=128} and \texttt{64} (146\(\rightarrow\)178\,ms), a symptom of batch decrease and stage-length skew—smaller effective batches amplify the sampling holdout, inflating the serial tail. 
\sysnames P99 progression (63\(\rightarrow\)82\(\rightarrow\)87\(\rightarrow\)36\,ms as rate decreases) remains smooth because the decision plane is \emph{stage-agnostic} and overlapped; residual variations reflect changes in batching efficiency rather than vocabulary-side stalls.

\subsection{Resource Utilization}
\label{sec:util}

We profile runtime resource usage of \sysname to confirm that \sysname sustains high GPU utilization with modest CPU/host memory overhead.

\textbf{GPU utilization.}
Figure~\ref{fig:watch-B200-gpu_util} reports B200 results (H100/L40 show the same trend in Appendix~\ref{appendix}). 
On B200, offloading and sequence-parallelizing the decision plane lifts mean GPU utilization from \(\mathbf{75\%}\) to \(\mathbf{96\%}\) (\(\mathbf{+21\%}\) on average; max \(\mathbf{+28\%}\) on \emph{Qwen3-235B-A22B}).

These improvements match our design: by removing sampling from the last PP stage and using asynchronous shared-memory streaming (\S\ref{sec:design}), the pipeline cycle \(1/T_{\text{cycle}}\) is no longer gated by a serial epilogue (Eq.~\ref{eq:cycle}).
With residency now in the mid-to-high 90s, the rate is set by non-sampling compute (GEMM/attention) rather than the decision plane.
Practically, steadier and higher GPU residency translates into more tokens/s at fixed latency and better scaling in \(t\) and \(p\); operators can meet the same load with fewer GPUs or tighten SLOs at similar cost.

\begin{figure}[t]
\centering
\includegraphics[width=\columnwidth]{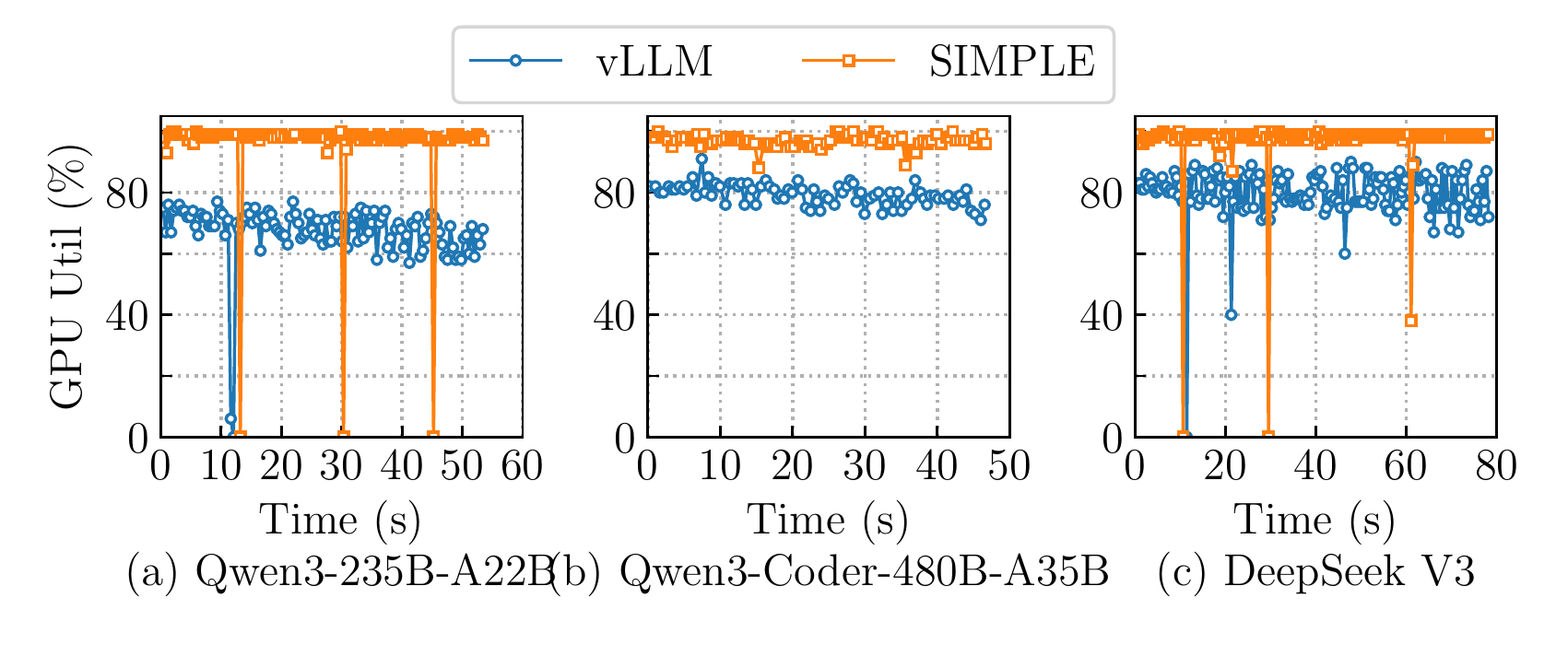}
\caption{B200: Runtime GPU utilization (mid-50\%) comparison (vLLM vs.\ \sysname).}
\label{fig:watch-B200-gpu_util}
\end{figure}

\begin{figure}[t]
  \centering
  \includegraphics[width=\columnwidth]{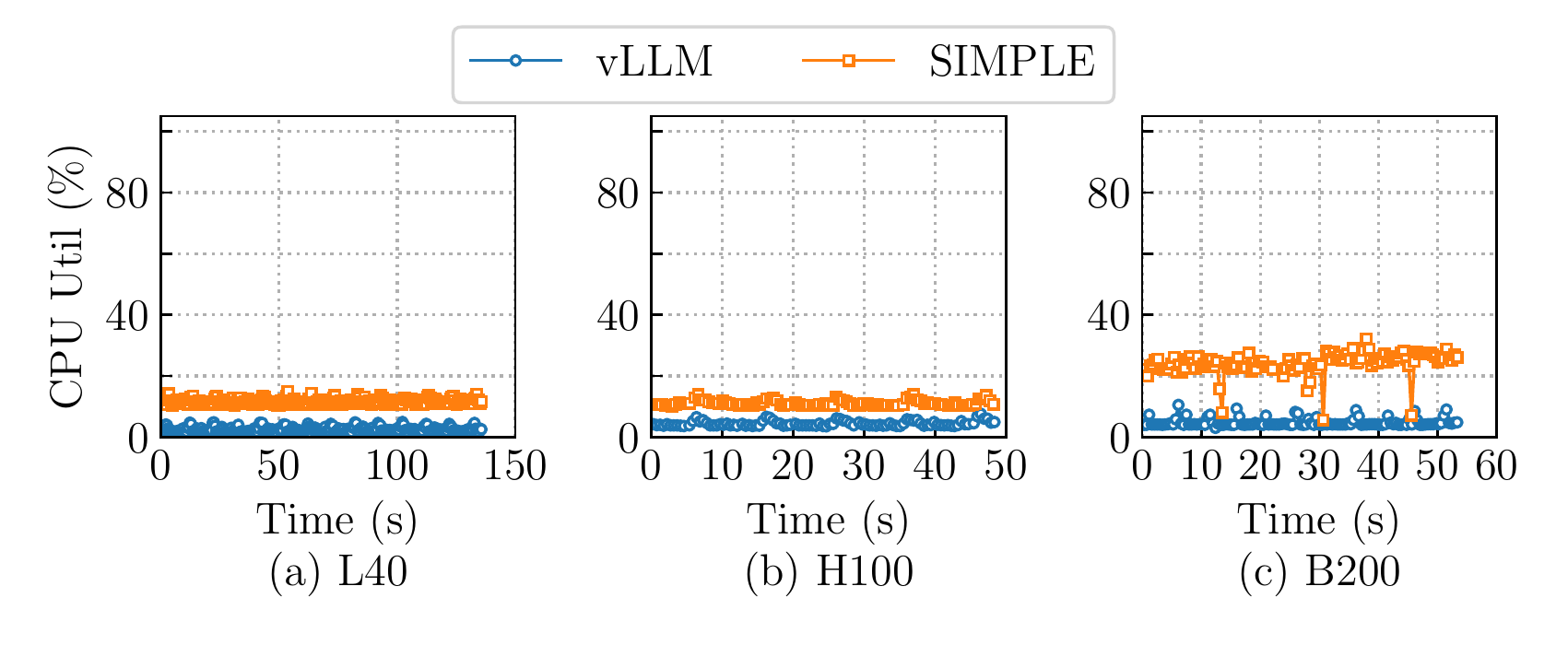}
  \caption{Runtime CPU utilization (mid-50\%) comparison (vLLM vs.\ \sysname) with \textbf{Qwen3-235B-A22B}.}
  \label{fig:watch-cpu_util}
\end{figure}

\textbf{CPU utilization.}
Offloading the decision plane increases CPU duty cycle as intended, but the magnitude varies by platform (Figure~\ref{fig:watch-cpu_util}).
On B200, the mean rise is \textbf{+17\%} across the three workloads; on L40, the mean rise is \textbf{+8\%}.
The gap stems from interconnect and compute throughput: B200’s NVLink and stronger kernels shorten \(T_{\text{stage},i}\) across the pipeline, so the decision plane must match a tighter \(T_{\text{cycle}}\) (Eq.~\ref{eq:cycle}); on L40 (PCIe~4.0), the GPU data plane is relatively slower, making the CPU work easier to fully overlap.

These patterns align with our single-pass, linear-time design and sizing model (\S\ref{sec:cpu}, \S\ref{sec:hotset-sizing}): CPU time scales with visited tokens and decode cadence, not directly with GPU FLOPs.
Despite higher CPU utilization on B200, utilization remains far from saturation (\(<\!31\%\) in our runs), confirming the decision plane stays \emph{overlappable} and off the critical path.
Practically, operators can allocate modest CPU headroom to unlock near-saturation GPU residency on modern nodes, while on PCIe-era nodes the CPU cost is even lower because overlap is easier.

\begin{table}[t]
  \centering
  \renewcommand{\arraystretch}{0.9}
  \caption{Host memory usage for Qwen3-235B-A22B.}
  \label{tab:host-mem-only}
  \begin{tabular}{lcc}
    \hline
    Platform & vLLM (\%) & SIMPLE (\%) \\
    \hline
    L40  & 3.9 & 4.6 \\
    H100 & 3.2 & 3.4 \\
    B200 & 6.8 & 8.1 \\
    \hline
  \end{tabular}
\end{table}

\textbf{Host memory usage.}
Across platforms on \textit{Qwen3-235B-A22B}, \sysname increases host memory utilization by at most \(\mathbf{+1.3\%}\) (6.8\%\(\rightarrow\)8.1\% on B200) with an average rise of \(\mathbf{+0.8\%}\), as shown in Table~\ref{tab:host-mem-only}.
This modest overhead is expected: shared-memory ring buffers are \emph{streamed} (not accumulated), and per-sampler state scales as \(O(B)+O(H)\). The column-wise layout and truncation-first filtering also reduce transient allocations, explaining the slight increase.
These results show that \sysname does not rely on multi-terabyte hosts; we expect the same mechanism to apply on more typical 256–512 GB inference nodes.

\subsection{Decision-Plane Microbenchmarks}
\label{sec:microbench}

\begin{figure}[t]
  \centering
  \includegraphics[width=\columnwidth]{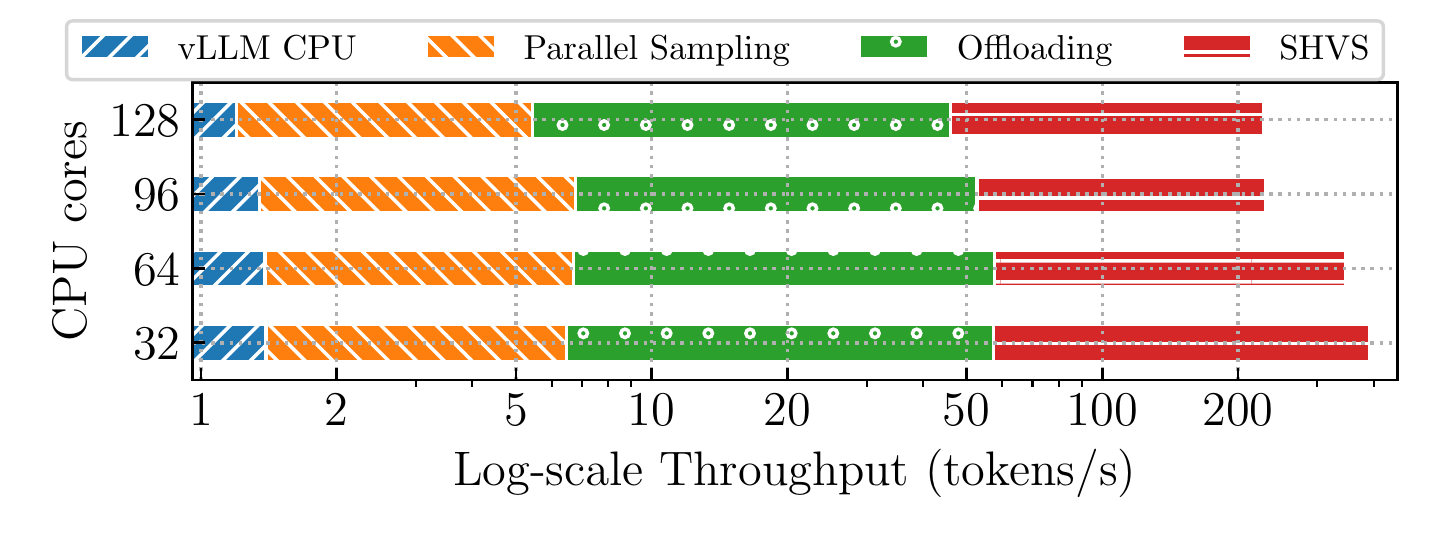}
  \caption{Per-sampler throughput (tokens/s, log-scale) of ablated designs on L40 with \textbf{QwQ-32B}.}
  \label{fig:marginal}
\end{figure}

\textbf{Ablations.}
To attribute gains and guide deployment, we isolate the decision plane and micro-benchmark per-sampler throughput (tokens/s) of \textit{QwQ-32B} under four variants: (i) \emph{vLLM CPU} (baseline full-\(V\) port), (ii) \emph{Parallel Sampling} (sequence-parallel but GPU-resident design), (iii) \emph{Offloading} (CPU-side, column-wise penalties, and truncation-first filtering), and (iv) \emph{SHVS} (hot-vocab with rejection sampling).
On average, per-sampler throughput rises from \textbf{1.3} (vLLM CPU) \(\rightarrow\) \textbf{6.4} (Parallel, \(+4.8\times\)) \(\rightarrow\) \textbf{53} (Offloading, \(+8.4\times\)) \(\rightarrow\) \textbf{300} (SHVS, \(+5.6\times\); \(+225\times\) over vLLM CPU in total).
At 32 threads, SHVS peaks at \textbf{393} tokens/s, illustrating the effect of cutting common-case work from \(O(V)\) to \(O(H)\) (\S\ref{sec:hotvocab}) on top of the single-pass CPU kernels (\S\ref{sec:cpu}).

All variants show mild per-sampler decline from 32\(\rightarrow\)128 threads (e.g., SHVS \(\,393\rightarrow228\) tokens/s).
This reflects \emph{shared-resource saturation} rather than algorithmic regressions:
more sampler threads compete for the same memory controllers and last-level cache, reducing effective bandwidth per thread.

These results indicate \sysnames decision plane is \emph{efficient} per sampler: modest CPU allocations suffice to keep GPUs near saturation (\S\ref{sec:util}). Practically, operators should \emph{right-size} the number of samplers \(m\) and the hot-vocab size \(H\) jointly—choose \(m\) to match the GPU cadence without hitting NUMA/bandwidth limits, and pick \(H\!\approx\!H^\star\) (\S\ref{sec:hotset-sizing}) to maximize sampler throughput.

\subsection{Hot-Vocab Sizing Model Validation}
\label{sec:sizing-validate}


\begin{figure}[t]
  \centering
  \subfigure[Cost fitting of SHVS.\label{fig:modeling:cost}]{
    \includegraphics[width=0.49\columnwidth]{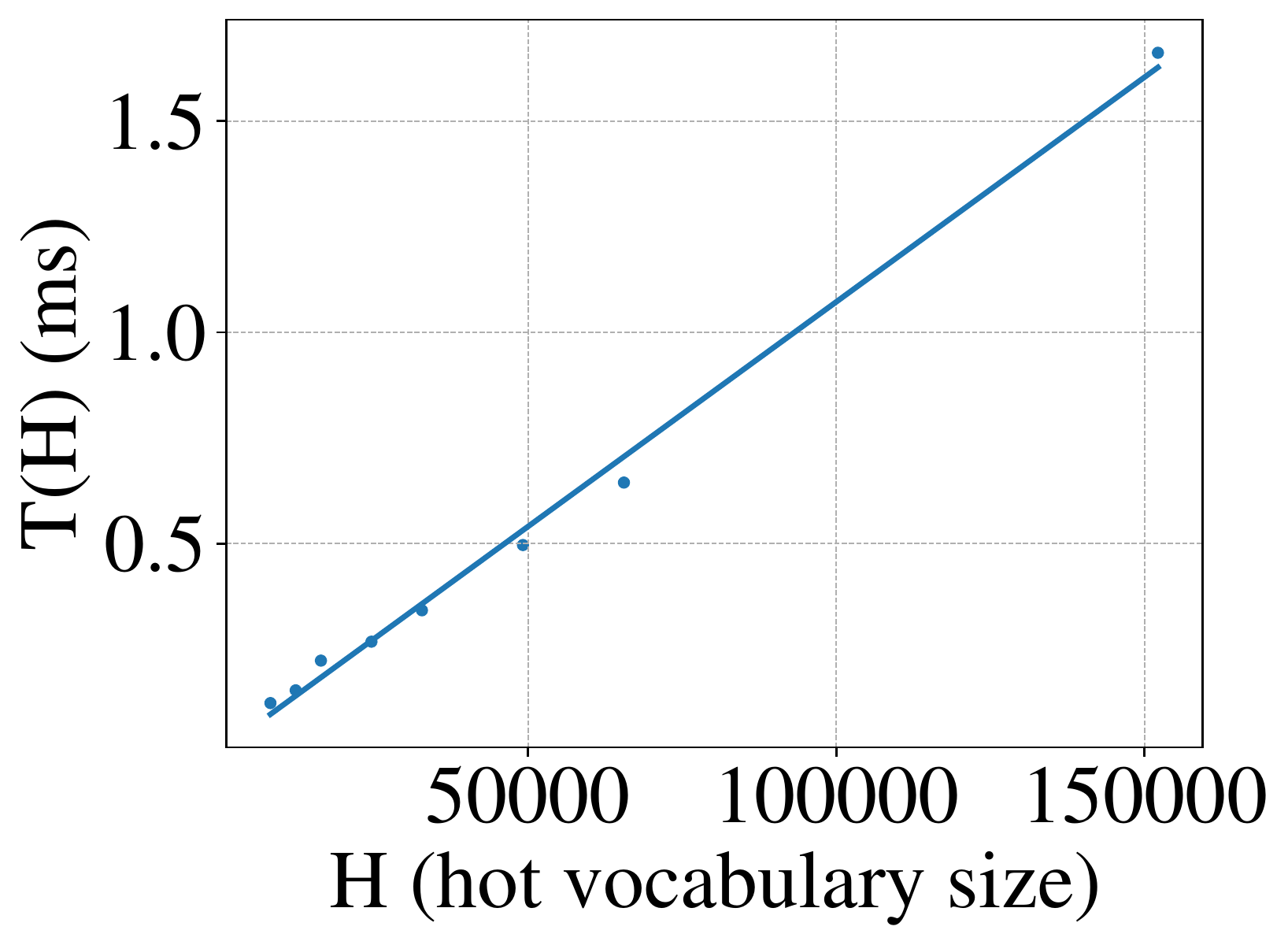}
  }\hfill
  \subfigure[Hit ratio of hot vocabulary with growing sizes.\label{fig:modeling:hit}]{
    \includegraphics[width=0.46\columnwidth]{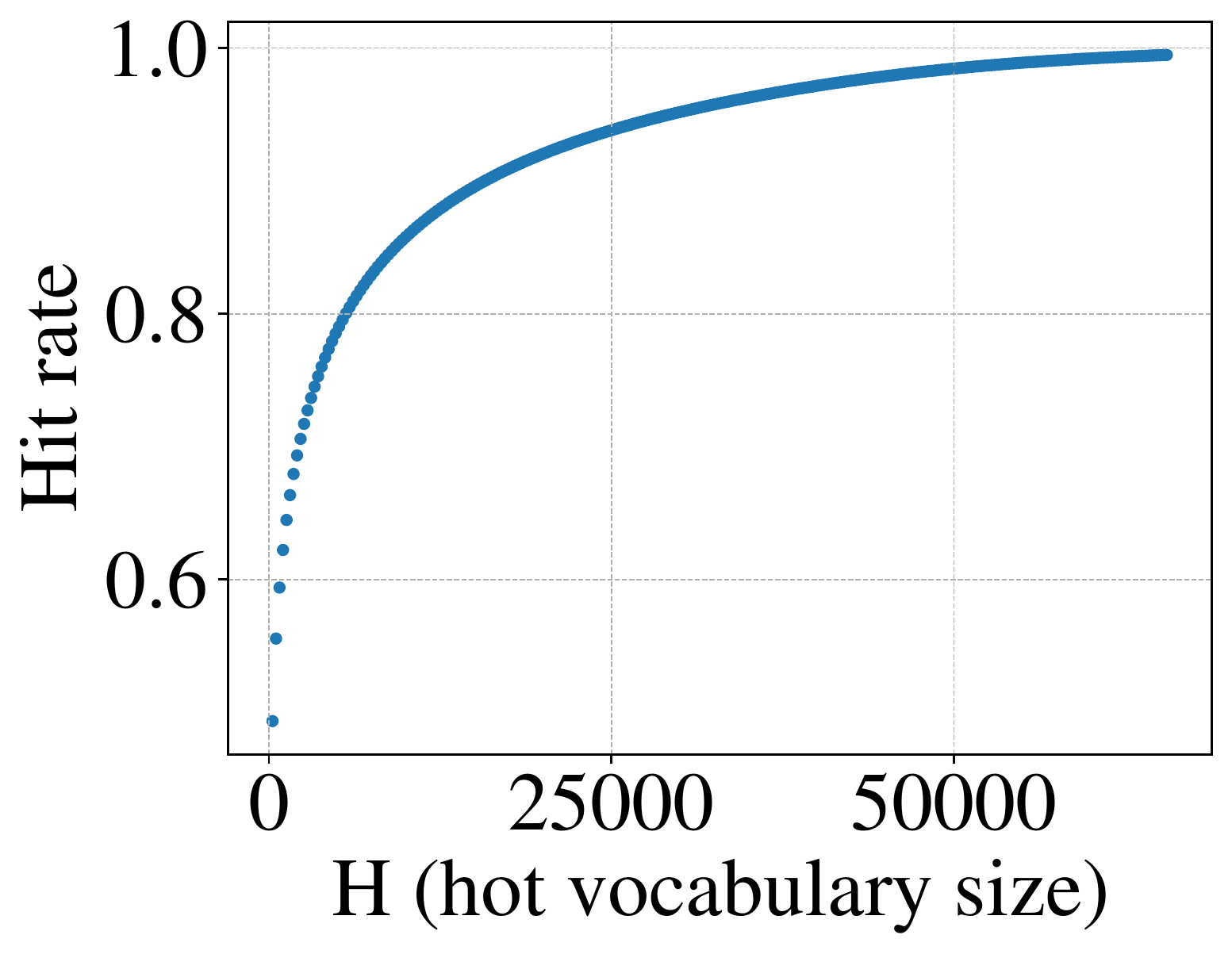}
  }
  \caption{Hot-vocab sizing modeling from measurements of \textbf{QwQ-32B} on L40.}
  \label{fig:modeling}
\end{figure}

\begin{figure}[t]
  \centering
  \subfigure[Optimizing target versus hot-vocab size.\label{fig:optimizing:objective}]{
    \includegraphics[width=0.45\columnwidth]{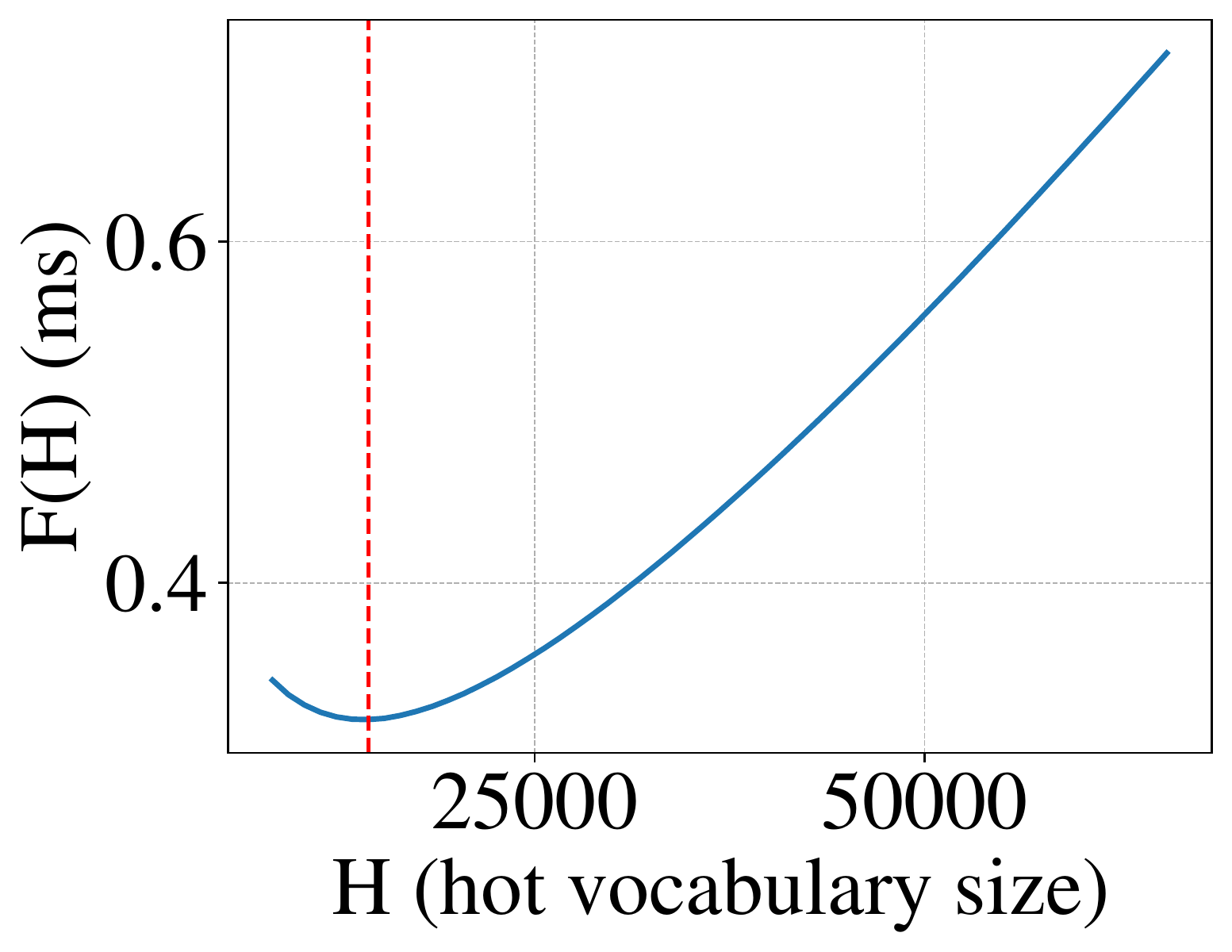}
  }\hfill
  \subfigure[Optimizing target versus real inference throughput.\label{fig:optimizing:fit}]{
    \includegraphics[width=0.5\columnwidth]{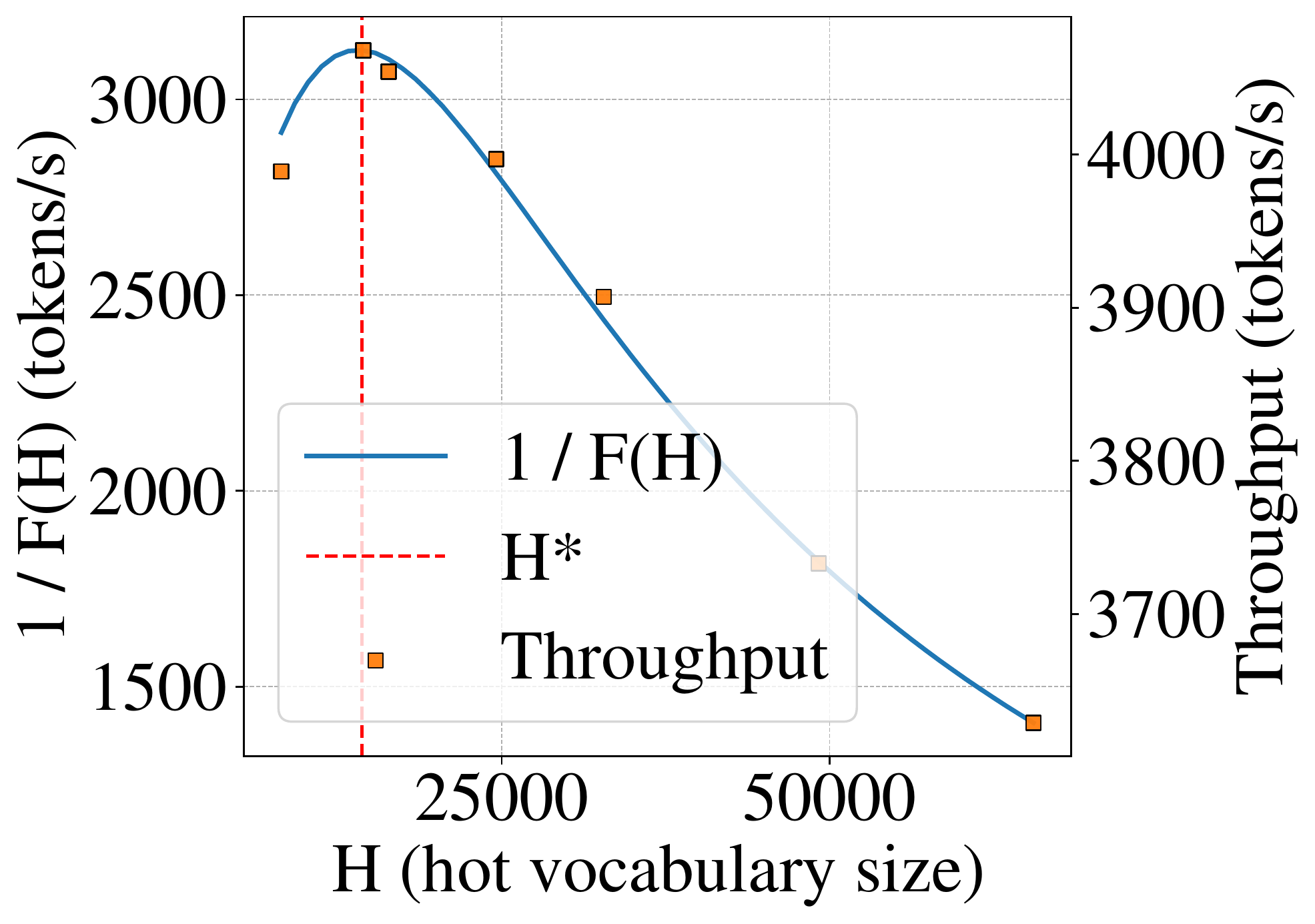}
  }
  \caption{Optimizing hot-vocab size for best performance with \textbf{QwQ-32B} on L40.}
  \label{fig:optimizing}
\end{figure}

We validate the sizing model from \S\ref{sec:hotset-sizing} using measurements on L40 with \textit{QwQ-32B}. Concretely, we fit the hot-path cost \(T_{\mathrm{cpu}}(H)\) by least squares, interpolate \(\bar{\alpha}(H)\) from traces, compose the expected decision cost
\(F(H)\) (cf.\ Eq.~\ref{eq:cpu-time}), and compare \(1/F(H)\) with real throughput. 

\textbf{Linear hot-path cost and hit ratio curve (Figure~\ref{fig:modeling}).}
The SHVS hot-path time fits an \emph{affine}, single-pass model \(T_{\mathrm{cpu}}(H)=c\,H+c_0\) with small residuals (\(c_0 = 8.55\times10^{-6}\), \(c = 1.06\times10^{-8}\) in Figure~\ref{fig:modeling:cost}), validating that CPU work grows linearly with the number of visited tokens in the hot set. This makes slope \(c\) (per-token scan cost) and intercept \(c_0\) (fixed per-sequence overhead) easy to estimate from a few points and underpins the affine form in Eq.~\eqref{eq:cpu-time}. 
In parallel, the hit-ratio curve \(\bar{\alpha}(H)\) is \emph{monotone} and \emph{saturating} (Figure~\ref{fig:modeling:hit}), consistent with Zipf-like mass concentration: small \(H\) already covers most probability, and marginal gains diminish as \(H\) grows. Because \(\bar{\alpha}(H)\) is model/policy–driven (hardware-agnostic), it can be profiled offline and reused across deployments.

\textbf{Expected cost and match to throughput (Figure~\ref{fig:optimizing}).}
Composing the two ingredients yields the expected decision cost \(F(H)\) (Eq.~\eqref{eq:cpu-time}), whose operating-region shape is convex-like with a single interior minimizer \(H^\star\) given in Eq.~\eqref{eq:stationary-h} (Figure~\ref{fig:optimizing:objective}). Intuitively, increasing \(H\) trades higher \(O(H)\) hot-path scans for a larger covered mass \(\bar{\alpha}(H)\) that avoids the \(O(V)\) tail; \(H^\star\) balances these forces and typically lies in a broad, forgiving valley. 
Overlaying \(1/F(H)\) with \emph{measured} tokens/s shows strong alignment in both the maximizer’s location and the surrounding shape (Figure~\ref{fig:optimizing:fit}): the predicted \(H^\star\) coincides with the empirical peak, and discrepancies only appear where non-sampling limits dominate. In practice, choosing \(H\!\approx\!H^\star\) and enforcing \(F(H)\!\le\!T_{\mathrm{cycle}}\) keeps the decision plane overlapped and maximizes end-to-end performance.

\subsection{Exactness of SHVS}
\label{sec:exactness}

\begin{figure}[t]
  \centering
  \includegraphics[width=\columnwidth]{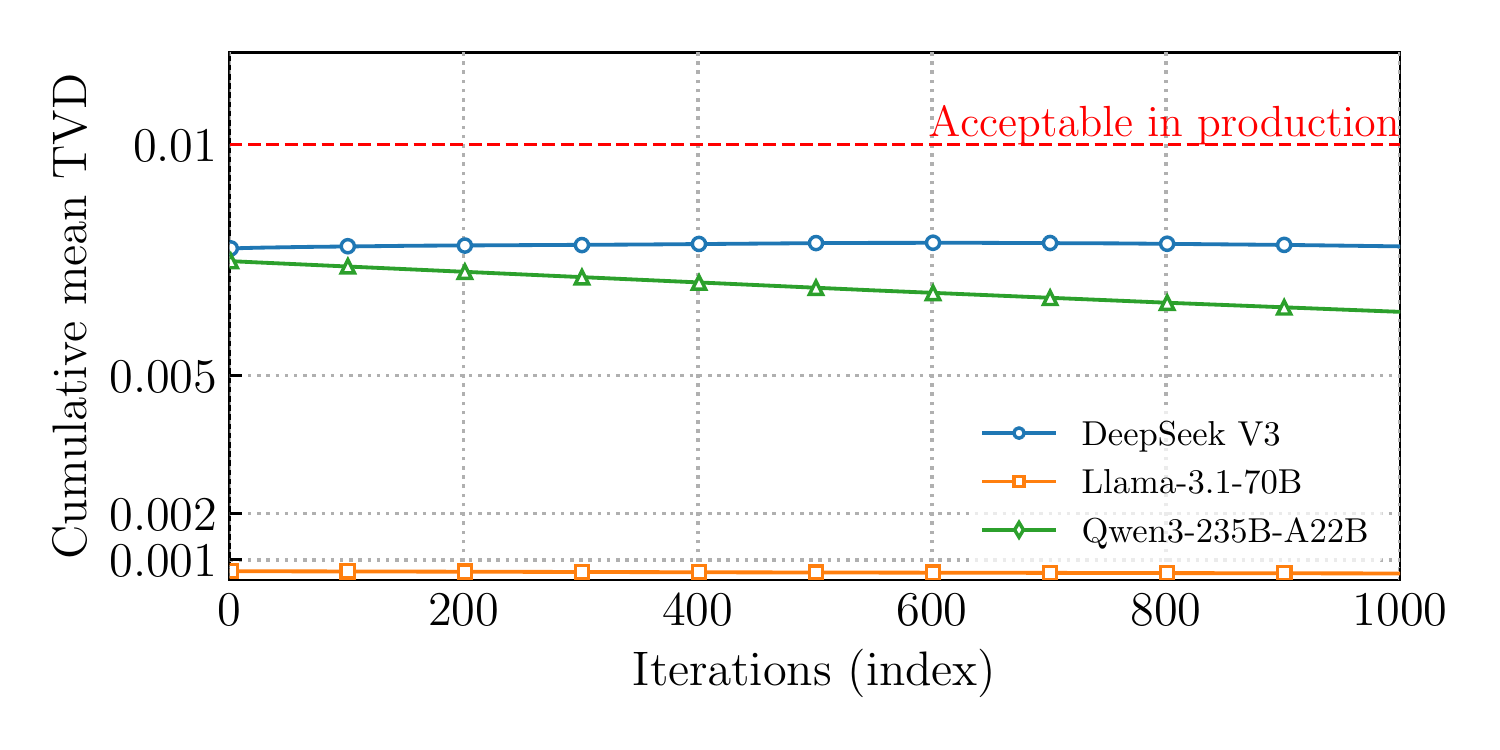}
  \caption{Cumulative mean TVD on H100 GPUs across models.}
  \label{fig:tvd}
\end{figure}

To empirically verify the distributional exactness guaranteed by Eq.~\ref{eq:hot-correctness}, we compute the total variation distance (TVD)~\cite{LevinPeresWilmer2009} between SHVS-generated next–token distributions and those from the baseline sampler at each decode step and for each sampler stream. We then plot the \emph{cumulative} TVD curves over the stable last 1K iterations for three models (DeepSeek~V3, Llama-3.1-70B, Qwen3-235B-A22B) on H100 GPUs, as shown in Figure~\ref{fig:tvd}.

Across all three models, the cumulative TVD curves are nearly flat with negligible slope and stay well below 1\% (e.g., \textbf{0.067\%} on average for Llama-3.1-70B), indicating that per-step TVD remains consistently low without drift. This behavior matches theory: SHVS’s accept/reject mechanism yields the same target distribution \(\tilde{p}\) (Eq.~\ref{eq:hot-correctness}); any residual TVD arises from finite-precision effects (e.g., temperature scaling, stable softmax) and stepwise changes in truncation support, not from bias in SHVS.

These results show that \sysnames performance gains (\S\ref{sec:hotvocab}, \S\ref{sec:hotset-sizing}) do not trade off sampling quality: the decision plane remains \emph{distributionally exact}. Practically, operators can tune \(H\) for throughput using the sizing model while maintaining output equivalence to a standard sampler.

\section{Related Work}
\label{sec:related}

\textbf{Model-parallel scaling and the data plane.}
FlashAttention makes exact attention I/O-aware and dramatically reduces HBM traffic~\cite{dao2022flashattention}. vLLM introduced PagedAttention and a high-throughput serving stack with efficient KV-cache management~\cite{kwon2023vllm}, while SGLang systematizes scheduling, batching, and runtime composition for LLM apps~\cite{zheng2024sglang}. These advances shrink GEMM/attention time but largely leave the \emph{decision plane} as a serial epilogue. As accelerators improve, this epilogue’s share grows (Amdahl’s law), becoming a scaling \emph{holdout}; \sysname directly targets this gap.

\textbf{Speculative decoding and drafting.}
Speculative decoding accelerates generation by proposing multiple tokens from a draft model and validating them with the target model~\cite{leviathan2023speculative, chen2023medusa}. These approaches reduce target-model steps but require extra models and validation passes. In contrast, \sysnames SHVS operates \emph{within} the decision plane of a single model: it samples on a small hot sub-vocabulary and applies rejection-correctness to preserve the exact distribution, avoiding auxiliary models and keeping GPU work unchanged.

\textbf{Disaggregation and service design.}
Recent work advocates disaggregation to rebalance resources and multiplex mixed workloads (e.g., prefill–decode separation, KV-centric serving)~\cite{zhong2024distserve, patel2024splitwise, hu2024tetrinfer, qin2025mooncake}. \sysname applies this principle specifically to the \emph{decision plane}: we externalize sampling as a CPU-side overlappable service. This design makes pipeline frequency no longer dictated by a serial epilogue and composes orthogonally with data-plane optimizations.
\section{Conclusion}
\label{sec:conclusion}
Sampling has become a structural \emph{holdout} in distributed LLM inference: it does not expand with tensor parallelism, extends the last pipeline stage in PP, and its share of iteration time grows as GPUs accelerate and vocabularies \(V\) grow. This paper introduces \sysname, a stage-agnostic, sequence-parallel, and overlappable decision-plane architecture that disaggregates sampling from the GPU data plane. 
Across L40/H100/B200 nodes and diverse models, \sysname delivers substantial end-to-end gains. 

\textit{Limitations and future work.} The decision plane still consumes CPU bandwidth; at very high thread counts, per-thread throughput can taper due to NUMA and memory-controller contention. When the hot-vocab mass \(\bar{\alpha}(H)\) is low (e.g., domain shift or heavy constraints), SHVS acceptance falls and benefits narrow; when the GPU data plane is compute-bound, headroom is limited. We therefore see opportunities for (i) online, QoS-aware controllers that adapt \(H\) using the sizing model; (ii) topology-aware placement to reduce NUMA traffic; and (iii) extending SHVS to structured/grammar-constrained decoding.

In sum, \sysname restores sampling to the inconspicuous role it should play: a small and largely hidden epilogue. By aligning the decision plane with modern \((t,p)\) scaling, it removes a stubborn serial tail and unlocks the next tranche of inference performance without user-side code changes.

\bibliography{reference}

@misc{shareGPT,
    author={ShareGPT},
    title = {{ShareGPT Datasets}},
    howpublished = {\url{https://huggingface.co/collections/bunnycore/sharegpt-datasets-66fa831dcee14c587f1e6d1c}},
    year = {2025}
}

@article{shoeybi2019megatron,
  title={Megatron-lm: Training multi-billion parameter language models using model parallelism},
  author={Shoeybi, Mohammad and Patwary, Mostofa and Puri, Raul and LeGresley, Patrick and Casper, Jared and Catanzaro, Bryan},
  journal={arXiv preprint arXiv:1909.08053},
  year={2019}
}

@misc{qwen2.5,
    title = {Qwen2.5: A Party of Foundation Models},
    url = {https://qwenlm.github.io/blog/qwen2.5/},
    author = {Qwen},
    month = {September},
    year = {2024}
}

@inproceedings{liu2024cachegen,
  title={Cachegen: Kv cache compression and streaming for fast large language model serving},
  author={Liu, Yuhan and Li, Hanchen and Cheng, Yihua and Ray, Siddhant and Huang, Yuyang and Zhang, Qizheng and Du, Kuntai and Yao, Jiayi and Lu, Shan and Ananthanarayanan, Ganesh and others},
  booktitle={Proceedings of the ACM SIGCOMM 2024 Conference},
  pages={38--56},
  year={2024}
}

@article{top-k,
  title={Hierarchical neural story generation},
  author={Fan, Angela and Lewis, Mike and Dauphin, Yann},
  journal={arXiv preprint arXiv:1805.04833},
  year={2018}
}

@article{top-p,
  title={The curious case of neural text degeneration},
  author={Holtzman, Ari and Buys, Jan and Du, Li and Forbes, Maxwell and Choi, Yejin},
  journal={arXiv preprint arXiv:1904.09751},
  year={2019}
}

@article{temperature-scaling,
  title={A learning algorithm for Boltzmann machines},
  author={Ackley, David H and Hinton, Geoffrey E and Sejnowski, Terrence J},
  journal={Cognitive science},
  volume={9},
  number={1},
  pages={147--169},
  year={1985},
  publisher={Elsevier}
}

@article{penalty-rep,
  title={Importance of a search strategy in neural dialogue modelling},
  author={Kulikov, Ilya and Miller, Alexander H and Cho, Kyunghyun and Weston, Jason},
  journal={arXiv preprint arXiv:1811.00907},
  volume={2},
  year={2018},
  publisher={Nov}
}

@misc{penalty-fre-pre,
    title = {{OpenAI API Documentation}},
    key = {penaltyfrepre},
    howpublished = {https://platform.openai.com/docs},
    author = {OpenAI},
    year = {2025}
}

@inproceedings{qin2025mooncake,
  title={Mooncake: Trading More Storage for Less Computation—A $\{$KVCache-centric$\}$ Architecture for Serving $\{$LLM$\}$ Chatbot},
  author={Qin, Ruoyu and Li, Zheming and He, Weiran and Cui, Jialei and Ren, Feng and Zhang, Mingxing and Wu, Yongwei and Zheng, Weimin and Xu, Xinran},
  booktitle={23rd USENIX Conference on File and Storage Technologies (FAST 25)},
  pages={155--170},
  year={2025}
}

@article{ouyang2022gpt,
  title={Training language models to follow instructions with human feedback},
  author={Ouyang, Long and Wu, Jeffrey and Jiang, Xu and Almeida, Diogo and Wainwright, Carroll and Mishkin, Pamela and Zhang, Chong and Agarwal, Sandhini and Slama, Katarina and Ray, Alex and others},
  journal={Advances in Neural Information Processing Systems},
  volume={35},
  pages={27730--27744},
  year={2022}
}

@article{brown2020gpt3,
  title={Language models are few-shot learners},
  author={Brown, Tom and Mann, Benjamin and Ryder, Nick and Subbiah, Melanie and Kaplan, Jared D and Dhariwal, Prafulla and Neelakantan, Arvind and Shyam, Pranav and Sastry, Girish and Askell, Amanda and others},
  journal={Advances in neural information processing systems},
  volume={33},
  pages={1877--1901},
  year={2020}
}

@misc{gptoss,
    key = {gptoss},
    title = {Introducing gpt-oss},
    note = {\url{https://openai.com/index/introducing-gpt-oss/}},
    year = {2025},
    author = {OpenAI}
}

@inproceedings{kwon2023vllm,
  title={Efficient memory management for large language model serving with pagedattention},
  author={Kwon, Woosuk and Li, Zhuohan and Zhuang, Siyuan and Sheng, Ying and Zheng, Lianmin and Yu, Cody Hao and Gonzalez, Joseph and Zhang, Hao and Stoica, Ion},
  booktitle={Proceedings of the 29th Symposium on Operating Systems Principles},
  pages={611--626},
  year={2023}
}

@inproceedings {zhong2024distserve,
author = {Yinmin Zhong and Shengyu Liu and Junda Chen and Jianbo Hu and Yibo Zhu and Xuanzhe Liu and Xin Jin and Hao Zhang},
title = {DistServe: Disaggregating Prefill and Decoding for Goodput-optimized Large Language Model Serving},
booktitle = {18th USENIX Symposium on Operating Systems Design and Implementation (OSDI 24)},
year = {2024},
isbn = {978-1-939133-40-3},
address = {Santa Clara, CA},
pages = {193--210},
url = {https://www.usenix.org/conference/osdi24/presentation/zhong-yinmin},
publisher = {USENIX Association},
month = jul
}

@article{cai2024pyramidkv,
  title={Pyramidkv: Dynamic kv cache compression based on pyramidal information funneling},
  author={Cai, Zefan and Zhang, Yichi and Gao, Bofei and Liu, Yuliang and Liu, Tianyu and Lu, Keming and Xiong, Wayne and Dong, Yue and Chang, Baobao and Hu, Junjie and others},
  journal={arXiv preprint arXiv:2406.02069},
  year={2024}
}

@article{team2025kimik2,
  title={Kimi K2: Open Agentic Intelligence},
  author={Team Kimi and Bai, Yifan and Bao, Yiping and Chen, Guanduo and Chen, Jiahao and Chen, Ningxin and Chen, Ruijue and Chen, Yanru and Chen, Yuankun and Chen, Yutian and others},
  journal={arXiv preprint arXiv:2507.20534},
  year={2025}
}

@article{xiao2023sink,
  title={Efficient streaming language models with attention sinks},
  author={Xiao, Guangxuan and Tian, Yuandong and Chen, Beidi and Han, Song and Lewis, Mike},
  journal={arXiv preprint arXiv:2309.17453},
  year={2023}
}

@article{li2024snapkv,
  title={Snapkv: {LLM} knows what you are looking for before generation},
  author={Li, Yuhong and Huang, Yingbing and Yang, Bowen and Venkitesh, Bharat and Locatelli, Acyr and Ye, Hanchen and Cai, Tianle and Lewis, Patrick and Chen, Deming},
  journal={arXiv preprint arXiv:2404.14469},
  year={2024}
}

@article{zheng2024sglang,
  title={Sglang: Efficient execution of structured language model programs},
  author={Zheng, Lianmin and Yin, Liangsheng and Xie, Zhiqiang and Sun, Chuyue and Huang, Jeff and Yu, Cody Hao and Cao, Shiyi and Kozyrakis, Christos and Stoica, Ion and Gonzalez, Joseph E and others},
  journal={arXiv preprint arXiv:2312.07104},
  year={2024}
}

@article{gage1994new,
  title={A new algorithm for data compression},
  author={Gage, Philip},
  journal={The C Users Journal},
  volume={12},
  number={2},
  pages={23--38},
  year={1994},
  publisher={R \& D Publications, Inc. Lawrence, KS, USA}
}

@inproceedings{kudo-richardson-2018-sentencepiece,
    title = "{S}entence{P}iece: A simple and language independent subword tokenizer and detokenizer for Neural Text Processing",
    author = "Kudo, Taku  and
      Richardson, John",
    editor = "Blanco, Eduardo  and
      Lu, Wei",
    booktitle = "Proceedings of the 2018 Conference on Empirical Methods in Natural Language Processing: System Demonstrations",
    month = nov,
    year = "2018",
    address = "Brussels, Belgium",
    publisher = "Association for Computational Linguistics",
    url = "https://aclanthology.org/D18-2012",
    doi = "10.18653/v1/D18-2012",
    pages = "66--71",
}

@inproceedings{sennrich-etal-2016-neural,
    title = "Neural Machine Translation of Rare Words with Subword Units",
    author = "Sennrich, Rico  and
      Haddow, Barry  and
      Birch, Alexandra",
    editor = "Erk, Katrin  and
      Smith, Noah A.",
    booktitle = "Proceedings of the 54th Annual Meeting of the Association for Computational Linguistics (Volume 1: Long Papers)",
    month = aug,
    year = "2016",
    address = "Berlin, Germany",
    publisher = "Association for Computational Linguistics",
    url = "https://aclanthology.org/P16-1162",
    doi = "10.18653/v1/P16-1162",
    pages = "1715--1725",
}

@misc{vllm_tunning,
    title = {{vLLM--Optimization and Tuning}},
    key = {vllm_tunning},
    author = {vLLM},
    howpublished = {https://docs.vllm.ai/en/latest/configuration/optimization.html},
    year = {2025}
}

@misc{vllm_ds3,
    title = {{Ray Serve DeepSeek with vLLM}},
    key = {vllm_ds3},
    author = {vLLM},
    howpublished = {\url{https://docs.vllm.ai/en/v0.8.5/getting_started/examples/ray_serve_deepseek.html}},
    year = {2025}
}

@misc{lambda_llama405,
    title = {Serving Llama 3.1 405B on a Lambda 1-Click Cluster},
    key = {lambda_405},
    howpublished = {\url{https://docs.lambda.ai/education/large-language-models/serving-llama-3-1-405b}},
    author = {Lambda Docs},
    year = {2025}
}

@misc{google_serve,
    title = {Google Kubernetes Engine (GKE)},
    key = {google_serve},
    howpublished = {\url{https://cloud.google.com/kubernetes-engine/docs/tutorials/serve-multihost-gpu}},
    author = {Google},
    year = {2025}
}

@article{zhang2023h2o,
  title={H2o: Heavy-hitter oracle for efficient generative inference of large language models},
  author={Zhang, Zhenyu and Sheng, Ying and Zhou, Tianyi and Chen, Tianlong and Zheng, Lianmin and Cai, Ruisi and Song, Zhao and Tian, Yuandong and R{\'e}, Christopher and Barrett, Clark and others},
  journal={Advances in Neural Information Processing Systems},
  volume={36},
  pages={34661--34710},
  year={2023}
}

@article{liu2024deepseekv3,
  title={Deepseek-v3 technical report},
  author={Liu, Aixin and Feng, Bei and Xue, Bing and Wang, Bingxuan and Wu, Bochao and Lu, Chengda and Zhao, Chenggang and Deng, Chengqi and Zhang, Chenyu and Ruan, Chong and others},
  journal={arXiv preprint arXiv:2412.19437},
  year={2024}
}

@article{guo2025deepseek,
  title={Deepseek-r1: Incentivizing reasoning capability in llms via reinforcement learning},
  author={Guo, Daya and Yang, Dejian and Zhang, Haowei and Song, Junxiao and Zhang, Ruoyu and Xu, Runxin and Zhu, Qihao and Ma, Shirong and Wang, Peiyi and Bi, Xiao and others},
  journal={arXiv preprint arXiv:2501.12948},
  year={2025}
}

@article{zhao2025fr,
  title={Fr-spec: Accelerating large-vocabulary language models via frequency-ranked speculative sampling},
  author={Zhao, Weilin and Pan, Tengyu and Han, Xu and Zhang, Yudi and Sun, Ao and Huang, Yuxiang and Zhang, Kaihuo and Zhao, Weilun and Li, Yuxuan and Wang, Jianyong and others},
  journal={arXiv preprint arXiv:2502.14856},
  year={2025}
}

@book{LevinPeresWilmer2009,
  title     = {Markov Chains and Mixing Times},
  author    = {Levin, David A. and Peres, Yuval and Wilmer, Elizabeth L.},
  year      = {2009},
  publisher = {American Mathematical Society},
  address   = {Providence, RI},
  isbn      = {978-0-8218-4739-8}
}

@misc{DeepSeek2025DeepGEMM,
  title     = {DeepGEMM: Clean and Efficient FP8/BF16 GEMM Kernels (incl. Grouped MoE)},
  author    = {{DeepSeek-AI}},
  year      = {2025},
  url       = {https://github.com/deepseek-ai/DeepGEMM},
  note      = {GPU GEMM library targeting LLM inference (FP8/BF16)}
}

@inproceedings{Elhoushi2025Any4,
  author    = {Mostafa Elhoushi and Jeff Johnson},
  title     = {any4: Learned 4-bit Numeric Representation for LLMs},
  booktitle = {Proceedings of the 42nd International Conference on Machine Learning (ICML '25)},
  year      = {2025},
  url       = {https://openreview.net/pdf?id=tJmhOPkWCj},
  note      = {Includes tinygemm: latency-optimized low-bit GEMM kernels}
}

@inproceedings{Lin2025QServe,
  author    = {Yujun Lin and Haotian Tang and Shang Yang and Zhekai Zhang and Guangxuan Xiao and Chuang Gan and Song Han},
  title     = {QServe: W4A8KV4 Quantization and System Co-design for Efficient LLM Serving},
  booktitle = {Proceedings of the 8th Conference on Machine Learning and Systems (MLSys '25)},
  year      = {2025},
  url       = {https://openreview.net/pdf/1ec600eaf0c56573a4d7a7818181657962d03d8f.pdf}
}

@inproceedings{dao2022flashattention,
  title={FlashAttention: Fast and Memory-Efficient Exact Attention with IO-Awareness},
  author={Dao, Tri and Fu, Dan and Ermon, Stefano and Rudra, Atri and R{\'e}, Christopher},
  booktitle={NeurIPS},
  year={2022}
}

@article{leviathan2023speculative,
  title={Fast Inference from Transformers via Speculative Decoding},
  author={Leviathan, Yaniv and Matias, Yossi and Polosukhin, Illia},
  journal={arXiv preprint arXiv:2211.17192},
  year={2023}
}

@article{chen2023medusa,
  title={Medusa: Simple LLM Inference Acceleration Framework with Multiple Draft Heads},
  author={Chen, Mingjie and Sun, Keqiang and others},
  journal={arXiv preprint arXiv:2309.11788},
  year={2023}
}

@inproceedings{patel2024splitwise,
  title={Splitwise: Efficient generative llm inference using phase splitting},
  author={Patel, Pratyush and Choukse, Esha and Zhang, Chaojie and Shah, Aashaka and Goiri, {\'I}{\~n}igo and Maleki, Saeed and Bianchini, Ricardo},
  booktitle={2024 ACM/IEEE 51st Annual International Symposium on Computer Architecture (ISCA)},
  pages={118--132},
  year={2024},
  organization={IEEE}
}

@article{hu2024tetrinfer,
  title={Inference without interference: Disaggregate llm inference for mixed downstream workloads},
  author={Hu, Cunchen and Huang, Heyang and Xu, Liangliang and Chen, Xusheng and Xu, Jiang and Chen, Shuang and Feng, Hao and Wang, Chenxi and Wang, Sa and Bao, Yungang and others},
  journal={arXiv preprint arXiv:2401.11181},
  year={2024}
}

@inproceedings{amdahl1967validity,
  author    = {Amdahl, Gene M.},
  title     = {Validity of the Single-Processor Approach to Achieving Large-Scale Computing Capabilities},
  booktitle = {Proceedings of the AFIPS Spring Joint Computer Conference},
  series    = {AFIPS '67 (Spring)},
  year      = {1967},
  pages     = {483--485},
  publisher = {ACM},
  address   = {New York, NY, USA},
  doi       = {10.1145/1465482.1465560}
}

@article{gustafson1988reevaluating,
  author  = {Gustafson, John L.},
  title   = {Reevaluating Amdahl's Law},
  journal = {Communications of the ACM},
  year    = {1988},
  volume  = {31},
  number  = {5},
  pages   = {532--533},
  doi     = {10.1145/42411.42415}
}

@article{little1961proof,
  author  = {Little, John D. C.},
  title   = {A Proof for the Queueing Formula: $L=\lambda W$},
  journal = {Operations Research},
  year    = {1961},
  volume  = {9},
  number  = {3},
  pages   = {383--387},
  doi     = {10.1287/opre.9.3.383}
}

@book{hintjens2013zeromq,
  title={ZeroMQ: messaging for many applications},
  author={Hintjens, Pieter},
  year={2013},
  publisher={" O'Reilly Media, Inc."}
}

@article{goel2025vocabtrim,
    title = "{VOCABTRIM}: Vocabulary Pruning for Efficient Speculative Decoding in {LLMs}",
    author = "Goel, Raghavv  and
              Agrawal, Sudhanshu  and
              Gagrani, Mukul  and
              Park, Junyoung  and
              Zao, Yifan  and
              Zhang, He  and
              Liu, Tian  and
              Yang, Yiping  and
              Yuan, Xin  and
              Lu, Jiuyan  and
              Lott, Chris  and
              Lee, Mingu",
    journal = "arXiv preprint arXiv:2506.22694",
    year = "2025",
    url = "https://arxiv.org/abs/2506.22694",
    note = "ICML 2025 Workshop on Efficient Systems for Foundational Models"
}

@article{zhang2025dynaspec,
    title = "{DynaSpec}: Context-aware Dynamic Speculative Sampling for Large-Vocabulary Language Models",
    author = "Zhang, Jinbin  and
              Ullah, Nasib  and
              Schultheis, Erik  and
              Babbar, Rohit",
    journal = "arXiv preprint arXiv:2510.13847",
    year = "2025",
    url = "https://arxiv.org/abs/2510.13847"
}

@misc{flashinfer_sampling_2025,
  title        = {Sorting-Free {GPU} Kernels for {LLM} Sampling},
  author       = {Xing, Shanli and Ye, Zihao and Hou, Bohan and Ceze, Luis and Chen, Tianqi},
  howpublished = {\url{https://flashinfer.ai/2025/03/10/sampling.html}},
  note         = {FlashInfer technical blog. Describes the Dual Pivot Rejection Sampling algorithm and fused GPU sampling kernels for top-k/top-p/min-p without full sorting},
  year         = {2025},
  month        = mar
}
\bibliographystyle{mlsys2025}




\end{document}